\documentclass{aastex63}
\usepackage{color}
\usepackage{CJK}

\submitjournal{ApJ}

\shorttitle{Two extremely low mass ratio contact binaries}
\shortauthors{Li et al.}

\begin{document}
\begin{CJK}{UTF8}{gbsn}

\title{Two contact binaries with mass ratios close to the minimum mass ratio}

\correspondingauthor{Kai Li}
\email{kaili@sdu.edu.cn}

\author{Kai Li}
\affil{Shandong Key Laboratory of Optical Astronomy and Solar-Terrestrial Environment, School of Space Science and Physics, Institute of Space Sciences, Shandong University, Weihai, Shandong, 264209, China}

\author{Qi-Qi Xia}
\affil{Shandong Key Laboratory of Optical Astronomy and Solar-Terrestrial Environment, School of Space Science and Physics, Institute of Space Sciences, Shandong University, Weihai, Shandong, 264209, China}

\author{Chun-Hwey Kim}
\affil{Department of Astronomy and Space Science, Chungbuk National University, Cheongju 361-763, Korea}

\author{Shao-Ming Hu}
\affil{Shandong Key Laboratory of Optical Astronomy and Solar-Terrestrial Environment, School of Space Science and Physics, Institute of Space Sciences, Shandong University, Weihai, Shandong, 264209, China}

\author{Di-Fu Guo}
\affil{Shandong Key Laboratory of Optical Astronomy and Solar-Terrestrial Environment, School of Space Science and Physics, Institute of Space Sciences, Shandong University, Weihai, Shandong, 264209, China}

\author{Min-Ji Jeong}
\affil{Department of Astronomy and Space Science, Chungbuk National University, Cheongju 361-763, Korea}

\author{Xu Chen}
\affil{Shandong Key Laboratory of Optical Astronomy and Solar-Terrestrial Environment, School of Space Science and Physics, Institute of Space Sciences, Shandong University, Weihai, Shandong, 264209, China}

\author{Dong-Yang Gao}
\affil{Shandong Key Laboratory of Optical Astronomy and Solar-Terrestrial Environment, School of Space Science and Physics, Institute of Space Sciences, Shandong University, Weihai, Shandong, 264209, China}

\begin{abstract}

The cut-off mass ratio is under debate for contact binaries. In this paper, we present the investigation of two contact binaries with mass ratios close to the low mass ratio limit. It is found that the mass ratios of VSX J082700.8+462850 (hereafter J082700) and 1SWASP J132829.37+555246.1 (hereafter J132829) are both less than 0.1 ($q\sim0.055$ for J082700, and $q\sim0.089$ for J132829). J082700 is a shallow contact binary with a contact degree of $\sim$19\%, and J132829 is a deep contact system with a fillout factor of $\sim$70\%. The $O-C$ diagram analysis indicated that both the two systems manifest long-term period decrease. In addition, J082700 exhibits a cyclic modulation which is more likely resulted from Applegate mechanism. In order to explore the properties of extremely low mass ratio contact binaries (ELMRCBs), we carried out a statistical analysis on contact binaries with mass ratios of $q\lesssim0.1$ and discovered that the values of $J_{spin}/J_{orb}$ of three systems are greater than 1/3. Two possible explanations can interpret this phenomenon. One is that some physical processes, unknown to date, are not considered when Hut presented the dynamically instability criterion. The other is that the dimensionless gyration radius ($k$) should be smaller than the value we used ($k^2=0.06$). We also found that the formation of ELMRCBs possibly has two channels. The study of evolutionary states of ELMRCBs reveals that their evolutionary states are similar with those of normal W UMa contact binaries.

\end{abstract}


\section{Introduction} \label{sec1}

Contact binaries which usually contain two late type Roche lobe filling components are very common in the field, open and globular clusters. It is estimated that about one of every 500 F, G, and K main sequence stars is a contact binary (\citealt{Rucinski2002}). Although contact binaries have been analyzed almost eighty years, there are still many issues to be settled, such as the formation, evolution and the ultimate fate (e.g., \citealt{Bradstreet1994,Eggleton2002,Qian2003,Yakut2005,Stepien2006}), the 0.22 days short period limit (e.g., \citealt{Rucinski1992,Stepien2006,Jiang2012,Qian2015,Li2019b}), the low mass ratio cut-off (e.g., \citealt{Rasio1995,Li2006,Arbutina2007,Arbutina2009,Jiang2010}), and the magnetic activities (e.g., \citealt{Applegate1992,Qian2007,Zhou2016a,Pi2017}). In order to solve these problems, we have to observe and study a large number of such systems.

The minimum mass ratio was first determined to be $q_{min}\sim0.09$ by \cite{Rasio1995} when neglecting the rotation of the less massive secondary component, and this value was dropped down to $q_{min}\sim0.076$ when considering the rotation of the secondary component and the dimensionless gyration radii $k^2_1=k^2_2=0.06$ (\citealt{Li2006}). \cite{Arbutina2007,Arbutina2009} derived the minimum value of mass ratio to be around $0.094\sim0.109$ by assuming a radiative primary and a fully convective secondary, and this value was decreased to be $0.070\sim0.074$ when taking into account the effect of ration. \cite{Jiang2010} suggested that the dimensionless gyration radii are decreasing with increasing mass and age, resulting in a minimum mass ratio different from 0.05 to 0.105. Very recently, based on a statistical study of 46 deep, low mass ratio contact binaries, \cite{Yang2015} determined a minimum mass ratio of about 0.044. Ultimately, such contact binaries are proposed to coalesce into a fast single rotation star, and it can be triggered not only when the spin angular momentum and orbital angular momentum meet the Darwin's instability (i.e. $J_{rot}>\frac{1}{3}J_{orb}$) but also when the contact degree exceeds 70\% or 86\% (\citealt{Hut1980,Eggleton2001,Rasio1995,Li2006}). \cite{Qian2005a} firstly put forward the concept "deep ($f \geq 50.0\%$), low mass ratio ($q \leq 0.25$) overcontact binaries" and proposed that this type of stars are likely to be the progenitors of blue stragglers/FK Com-type stars. At present, a lot of such type binaries have been analyzed (e.g., \citealt{Yang2009,Qian2011,Liao2017}).
However, there is only one target, V1309 Sco, whose merging progress has been observed (\citealt{Tylenda2011}), and it is making a blue straggler (\citealt{Ferreira2019}). \cite{Zhu2016} derived that V1309 Sco is a very deep contact binary ($f=89.5\%$) with extremely low mass ratio ($q\sim0.095$) before the merge. Therefore, in order to comprehend the low mass ratio cut-off and search for progenitors of the merger, we should observe and analyze more contact binaries with mass ratios less than 0.1. In this paper, we present the observations and investigations of two such systems, J082700 and J132829.

J082700 was firstly classified as a W UMa type star by G. Srdoc in 2010 (The International Variable Star Index, VSX\footnote{https://www.aavso.org/vsx/index.php?view=detail.top\&oid=251475}). The period and amplitude of the light variation were determined to be 0.2717 days and 0.157 mag in the V bandpass. J132829 was firstly classified as a W UMa type star by T\'{e}lescope \`{a} Action Rapide pour les Objets Transitoires (TAROT, \citealt{Damerdji2007}), and the period was derived to be 0.384718 days. Recently, the orbital period and amplitude of the light variation were updated to be 0.2771582 days and 0.14 mag in the V bandpass for J082700 and 0.3847052 days and 0.13 mag in the V bandpass for J132829 by the All-Sky Automated Survey for SuperNovae (ASAS-SN, \citealt{Shappee2014,Jayasinghe2018}). In this paper, the multiple color light curves of the two systems were presented, then their orbital period changes were studied, and a statistic on contact binaries with mass ratios $q\lesssim0.1$ is shown.

\section{Observations and Data Reduction}
The multiple color light curves of J082700 and J132829 were both observed in 2019 by using one-meter class telescopes. J082700 was observed using $VR_cI_c$ filters by the Weihai Observatory 1.0-m telescope of Shandong University (WHOT, \citealt{Hu2014}) on March 23, 25, and 31 (the seeing was around $1.7^{\prime\prime}$). J132829 was observed using $BVR_cI_c$ filters by 85 cm telescope at the Xinglong Station of National Astronomical Observatories (NAOs85cm) in China on March 15 (the seeing was around $4.5^{\prime\prime}$), and by WHOT on April 25 and May 2 (the seeing was around $2.0^{\prime\prime}$). During the observations, the PIXIS 2048B CCD camera and the Andor DZ936 CCD camera were equipped on WHOT and NAOs85cm, respectively. The scale of each image is approximately 0.35$\arcsec$ per pixel for WHOT, resulting in a field of view of about 12$'$ $\times$ 12$'$, and the scale of each image is approximately 0.94$\arcsec$ per pixel for NAOs85cm, resulting in a field of view of about 32$'$ $\times$ 32$'$. During the observations, J082700 was exposed for 80, 50 and 30 seconds in the $V$, $R_c$ and $I_c$ bands, respectively. The exposure time of J132829 was 50s, 30s, 20s, and 15s  in the $B$, $V$, $R_c$ and $I_c$ bands, respectively, when using NAOs85cm, while 70s, 40s, 25s, and 20s in the $B$, $V$, $R_c$ and $I_c$ bands, respectively, when using WHOT. All images of the two stars were reduced by using the IRAF package. During the reduction, the comparison and check stars are GSC 3415-2327 ($J=12.669$ mag, $H=12.209$ mag, $K=12.102$ mag) and GSC 3415-2313 ($J=12.852$ mag, $H=12.583$ mag, $K=12.520$ mag) for J082700, while those for J132829 are GSC 3853-0012 ($J=10.335$ mag, $H=9.884$ mag, $K=9.828$ mag) and GSC 3853-0279 ($J=10.616$ mag, $H=10.226$ mag, $K=10.156$ mag).

The aperture photometry method was applied to reduce the observational data of J082700. When reducing the NAOs85cm data of J132829, we used the PSF photometry method because there is a very close companion star (2MASS J13282957+5552471) which is about $7^{\prime\prime}$ from J132829 and the seeing during the observations was around $4.5^{\prime\prime}$ which is more than a half of the distance between the two stars. While for the WHOT data, we found that the photometric results derived by the aperture photometry method are much better than those derived by the PSF photometry method (the seeing during the observations was around $2.0^{\prime\prime}$ which is less than one third of the distance), so we used the aperture photometry method to reduce the WHOT data of J132829.
After the reduction, we determined one set of complete light curve for J082700 and two sets of complete light curve for J132829. All the light curves are shown in Figure 1. The photometric accuracy for the light curve of J082700 is around 0.006 mag in $V$ band, 0.005 mag in $R_c$ band, and 0.008 mag in $I_c$ band, that for NAOs85cm light curve of J132829 is around 0.009 mag in $B$ band, 0.007 mag in $V$ band, 0.008 mag in $R_c$ band, and 0.007 mag in $I_c$ band, and that for WHOT light curve of J132829 is around 0.006 mag in $B$ band, 0.005 mag in $V$ band, 0.005 mag in $R_c$ band, and 0.007 mag in $I_c$ band. Due to the light curve of J082700, we found that the amplitude of light variation in V band is definitely 0.14 mag. Due to the light curves of J132829, the V band amplitude of light variability is about 0.29 mag.  The 0.13 mag light variability determined by ASAS-SN (\citealt{Shappee2014,Jayasinghe2018}) should be caused by the large pixel scale of the camera which is about $8^{\prime\prime}$, thus the ASAS-SN camera cannot separate the target and the companion star. From our observations, we determined one heliocentric time of light minimum of J082700 to be $2458574.10494\pm0.00055$ and three heliocentric times of light minimum of J132829 to be $2458558.04146\pm0.00089$, $2458558.23706\pm0.00088$, and $2458599.20602\pm0.00036$. The method of \cite{Kwee1956} (hereafter K-W) was used to determine those times of minima. As seen in Figure 1, the two binaries are both totally eclipsing binaries.

\begin{figure*}
\begin{center}
\includegraphics[width=0.33\textwidth]{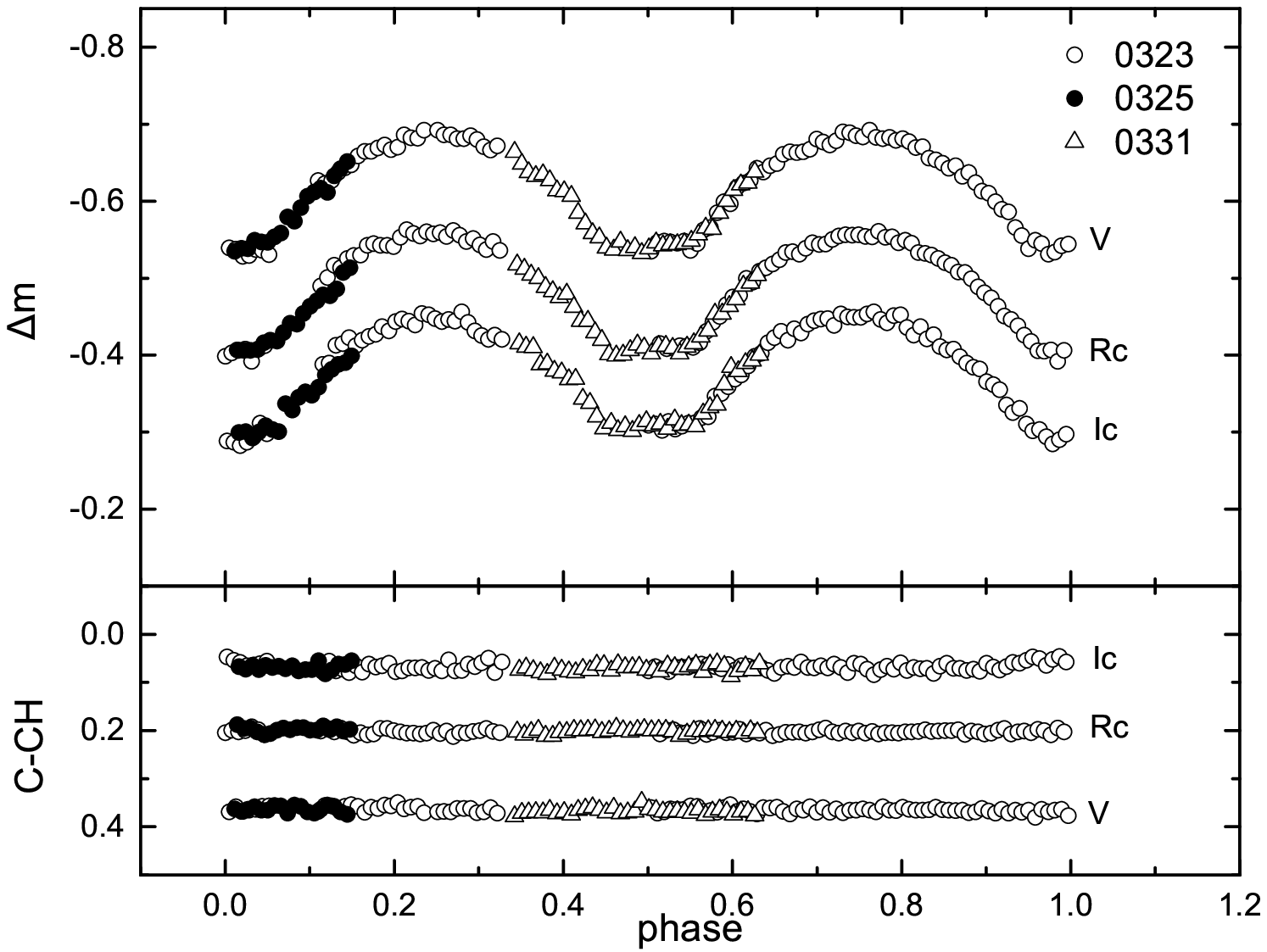}%
\includegraphics[width=0.32\textwidth]{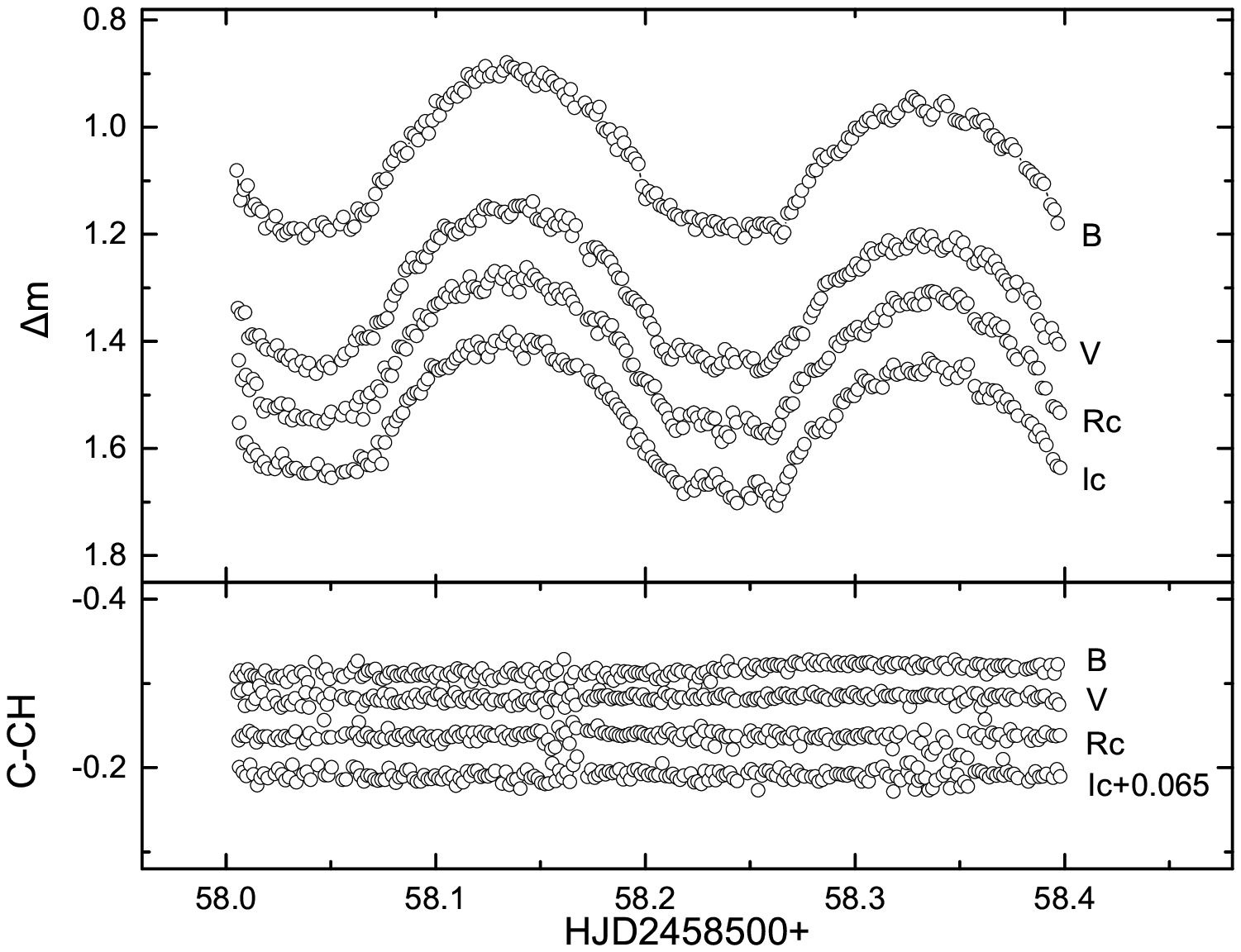}
\includegraphics[width=0.33\textwidth]{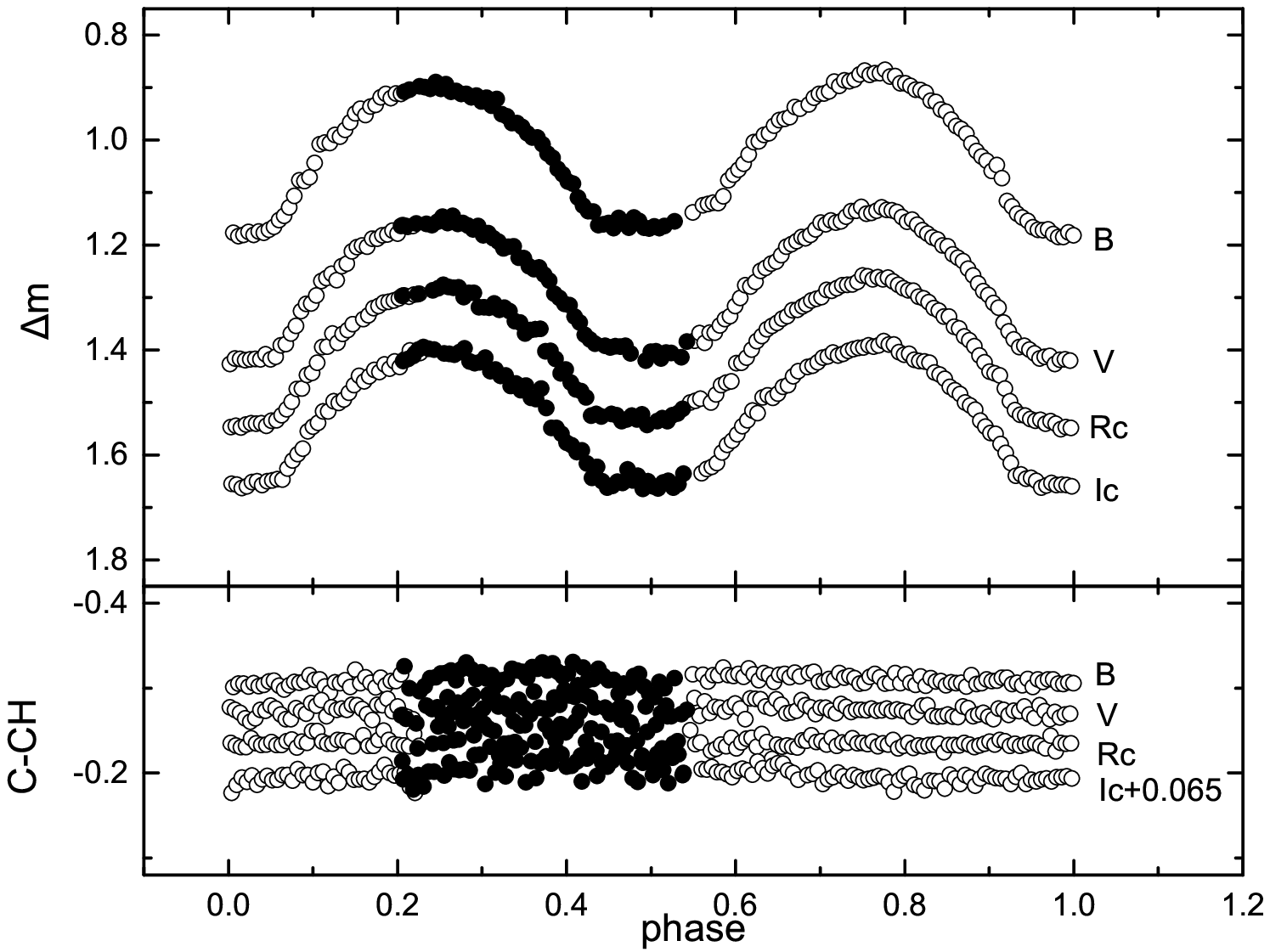}

\caption{The left panel displays the $VR_cI_c$ light curves of J082700 (the open, solid circles, and open triangles respectively refer to the data observed on March 23, 25, and 31, 2019), the middle panel plots the $BVR_cI_c$ light curves observed by NAOs85cm of J132829, while the right one plots the $BVR_cI_c$ light curves observed by WHOT of J132829 (the open and solid circles respectively represent the data observed on April 25 and May 2, 2019).}
\end{center}
\end{figure*}

\section{Light Curve Solution}
In order to determine the photometric elements of the two binaries, we should determine the effective temperature at first. Thanks to the Gaia mission (\citealt{Gaia2016,Gaia2018}), the temperatures have been determined to be 5847 K for J082700 and 6329 K for J132829, respectively. According to \cite{Andrae2018}, the typical accuracy for the effective temperatures derived by Gaia between 3000 and 10000 K is estimated to be $\pm324$ K.
The gravity darkening and bolometric albedo coefficients of the two components are both set as $g_1=g_2=0.32$ and $A_1=A_2=0.5$ due to \cite{Lucy1967} and \cite{Rucinski1969}. The atmospheric models of \citep{Castelli2004} were applied and the limb-darkening coefficients were derived by a logarithmic law.

The mass ratio is one of the most important parameters for contact binaries and can be well determined by radial velocity curve. If there is no radial velocity curve, the mass ratio can be well determined by only using photometric light curves for totally eclipsing contact binaries because they have very steep mass ratio relative radii relationship (\citealt{Terrell2005}). Both of our two targets manifest totally eclipsing phenomenon, indicating we can obtain accurate mass ratios.
The Markov Chain Monte Carlo (MCMC) parameter search method by using PHOEBE 2.3 \citep{Prsa2005,Prsa2016,Horvat2018,Jones2020,Conroy2020} under contact binary mode was employed to determine the most probabilistic physical parameters of the two binaries. The python-based MCMC software EMCEE \citep{Foreman2013} was used to adjust the parameters $q$, $i$, $T_2$, $r_1$, and $l_2$ for J082700 and $q$, $i$, $T_2$, $r_1$, $l_2$, $\theta$, $\lambda$, $r_s$, $T_s$, and $l_3$ for J132829. The values of these parameters determined by W-D code \citep{Wilson1971,Wilson1979,Wilson1990} were served as priors for the MCMC sampling. A total of 32 parameter space "Walkers" were applied for the two binaries. Because the light curves shown in Figure 1 are obvious deviations from an "ideal" light curve shape in pretty much all passbands, which inevitably means that the employed model (in this case PHOEBE) will suffer from systematics, we applied gaussian processes during the modelling procedure. If we assigned the Gaia temperature to the primary component, implicitly assuming that the surface brightness of the secondary component is negligible. This is only approximate, the surface brightness of the secondary star cannot be negligible. In order to solve this problem, we assumed blackbody radiation and used the following equations to determine the individual temperatures \citep{Zwitter2003,Christopoulou2013},
\begin{eqnarray}
T_1&=&(((1+k^2)T_{eff}^4)/(1+k^2(T_2/T_1)^4))^{0.25},  \\\nonumber
T_2&=&T_1(T_2/T_1),
\end{eqnarray}
where $T_{eff}$ is the temperature determined by Gaia, $k$ is the radius ratio, and $T_2/T_1$ is the temperature ratio. However, the surface brightness ratio does not scale linearly with the temperature ratio. Therefore, we carried out three iterations to determine a relatively more precise primary temperature. During each iteration,
the MCMC parameter search was run 2000 steps for all the three sets of light curves, resulting 64000 iterations (the burn-in part has been deleted when calculating the physical parameters). ¡°Chains longer than 10 times of the integrated autocorrelation time¡± was chosen as the convergent criterion following \cite{Conroy2020}. We found that the chains are longer than 20 times of the integrated autocorrelation time for each iteration, indicating all the MCMC runs are convergent. Each iteration needs the primary temperature as an input parameter. Firstly, the Gaia temperature was assigned to the primary, and a more precise primary temperature was calculated by Equation (1). Secondly, a new iteration was carried out using the more precise primary temperature. Thirdly, new primary temperature was obtained by Equation (1) and was applied for the third iteration. We found that the physical parameters change very little during the three iterations, especially the primary temperature, the differences are less than 15 K for the three sets of light curves (5855 K, 5862 K, and 5867 K for J082700, 6313 K, 6299 K, and 6287 K for the NAOs85cm light curves of J132829, and 6315 K, 6305 K, and 6302 K for the WHOT light curves of J132829 for the three iterations). After that, 5867 K was assigned to the primary temperature of J082700, 6287 K was assigned to the primary temperature of the NAOs85cm light curves of J132829, and 6302 K was assigned to the primary temperature of the WHOT light curves of J132829. New MCMC parameter searching was carried out with a step of 15000, resulting 480000 iterations. This time, the chains are longer than 60 times of the integrated autocorrelation time, indicating very good convergence. When we have finished the MCMC parameter searching, the first 160000 iterations were discarded.
Figures 2-4 shows the probability distributions of $q$, $i$, $f$, and $l_2/l_1$ for J082700 and the NAOs85cm light curves of J132829 and the WHOT light curves of J132829, respectively. All the physical parameters were obtained by using the median-value and are listed in Table 1. The final individual temperatures of the two components were also calculated by Equation (1) and are listed in Table 1, the uncertainty of the Gaia temperature was taken into account when estimating their uncertainties.
For J132829, we found that the results derived by NAOs85cm light curves and those derived by the WHOT light curves are consistent with each other. The photometric elements determined by the WHOT light curves were used for the following analysis because of the higher precision.
The synthetic light curves calculated by PHOEBE 2.3 are all plotted in Figure 5. The left panel of this figure displays the fitted light curves of J082700, the middle panel plots the fitted NAOs85cm light curves of J132829, while the right one plots the fitted WHOT light curves of J132829. The O-C residuals are plotted in the bottom panels, nearly flat residuals reveal that the theoretical light curves fit the observed ones very well.

\begin{figure*}
\centering
\includegraphics[width=16cm, angle=0]{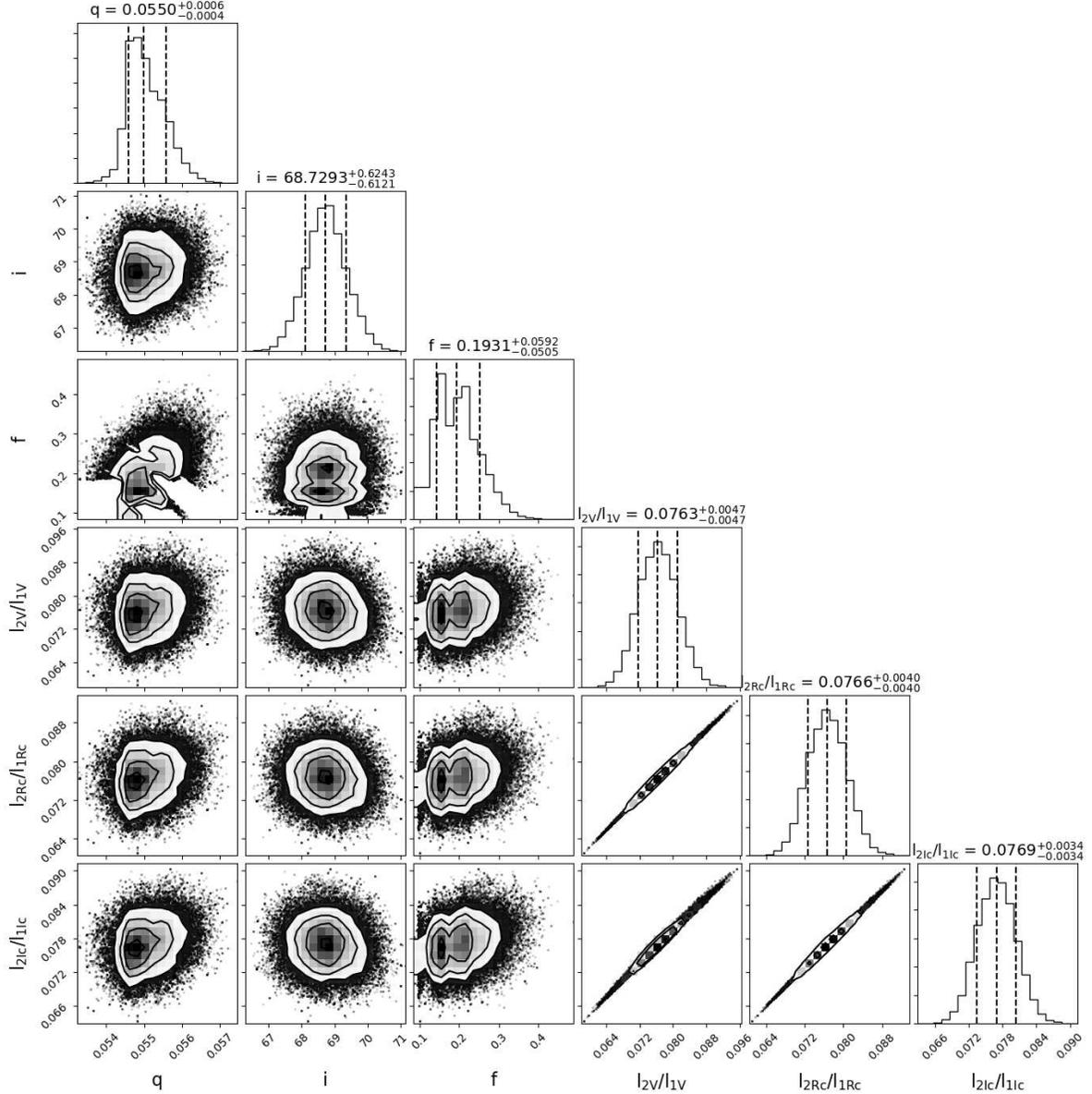}
\caption{The probability distributions of $q$, $i$, $f$, and $l_2/l_1$ determined by the MCMC modeling for J082700.}
\end{figure*}

\begin{figure*}
\centering
\includegraphics[width=16cm, angle=0]{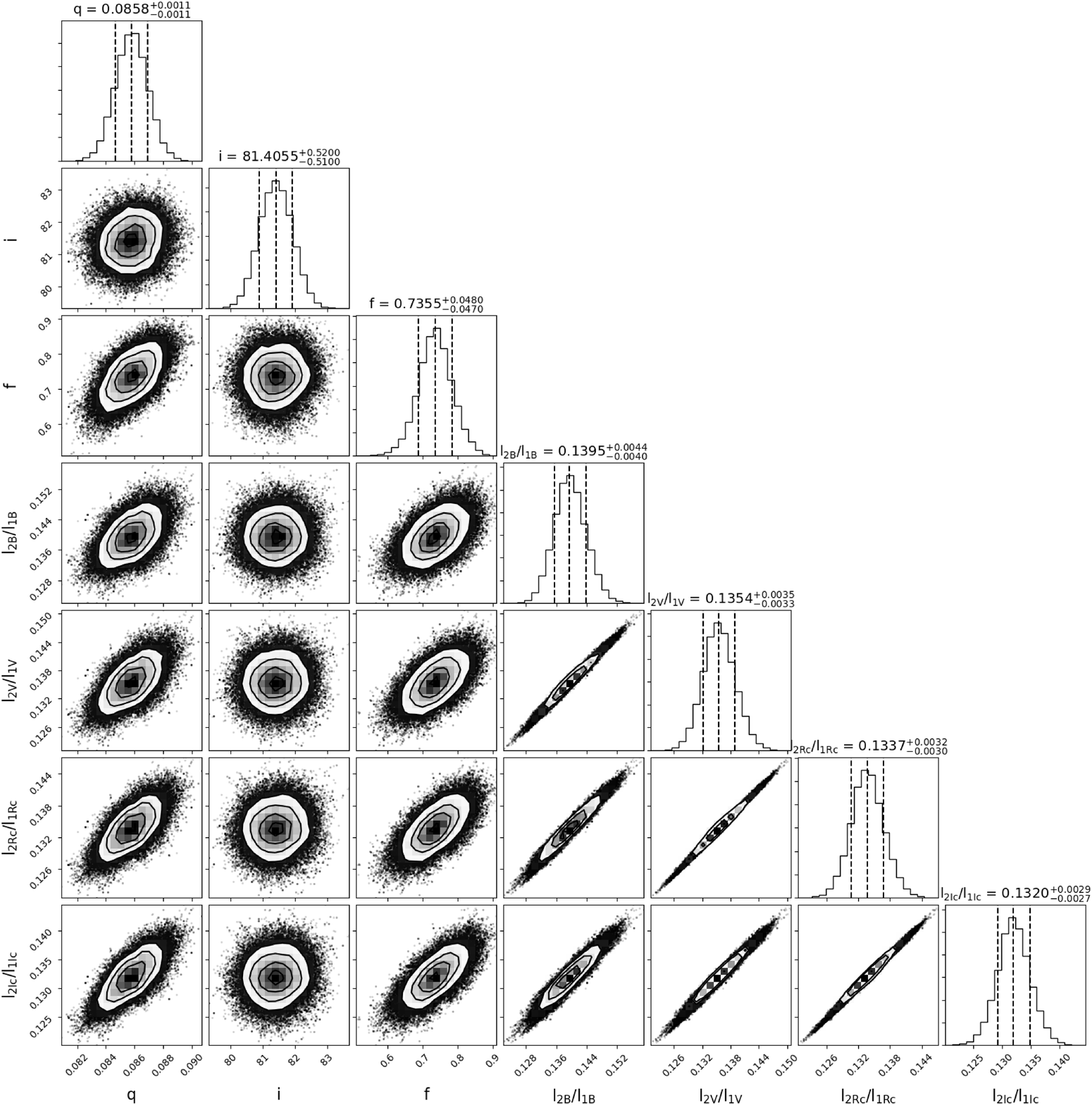}
\caption{The probability distributions of $q$, $i$, $f$, and $l_2/l_1$ determined by the MCMC modeling for the NAOs85cm light curves of J132829.}
\end{figure*}

\begin{figure*}
\centering
\includegraphics[width=16cm, angle=0]{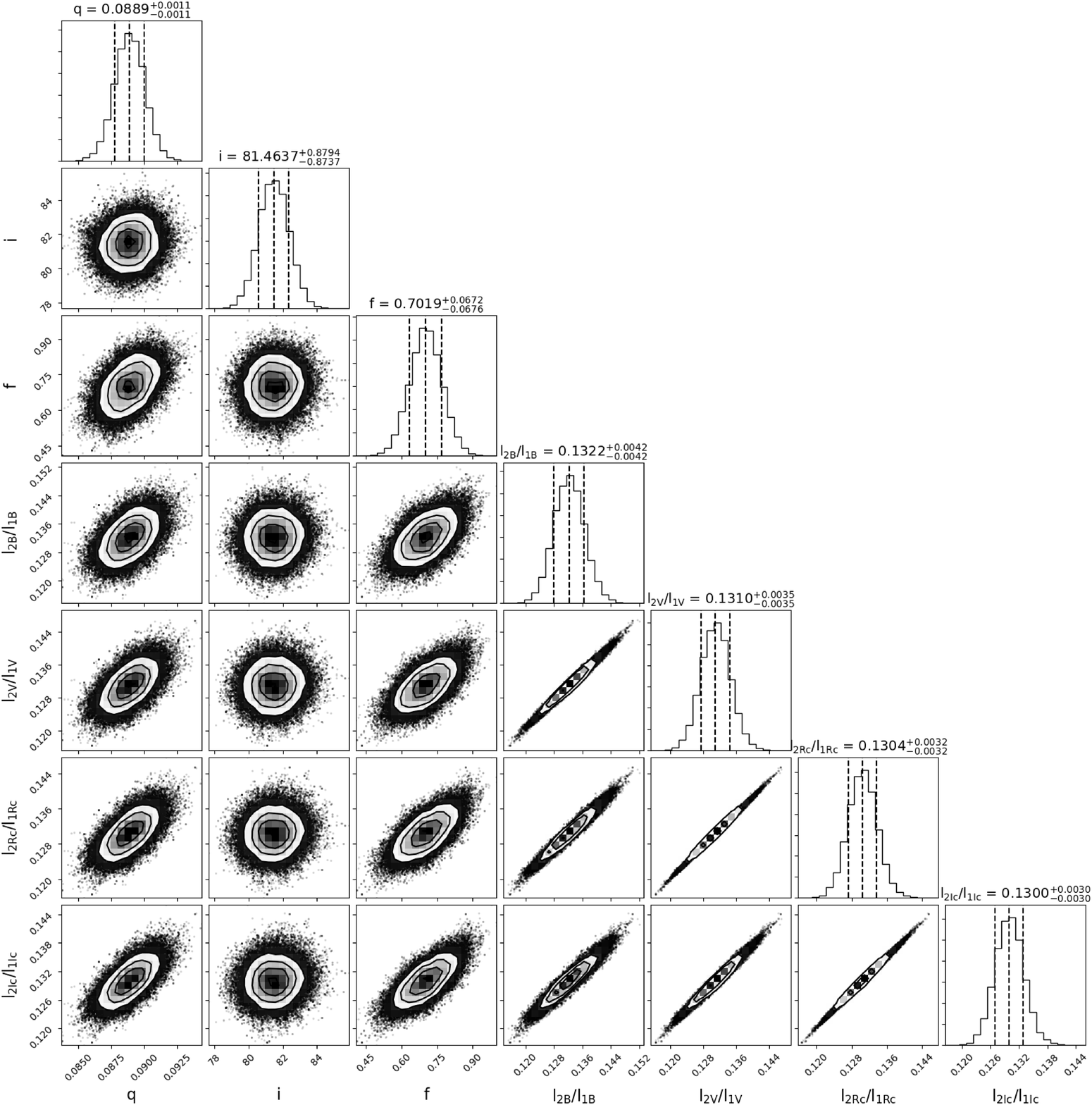}
\caption{The probability distributions of $q$, $i$, $f$, and $l_2/l_1$ determined by the MCMC modeling for the WHOT light curves of J132829.}
\end{figure*}

\begin{figure*}
\begin{center}
\includegraphics[width=0.325\textwidth]{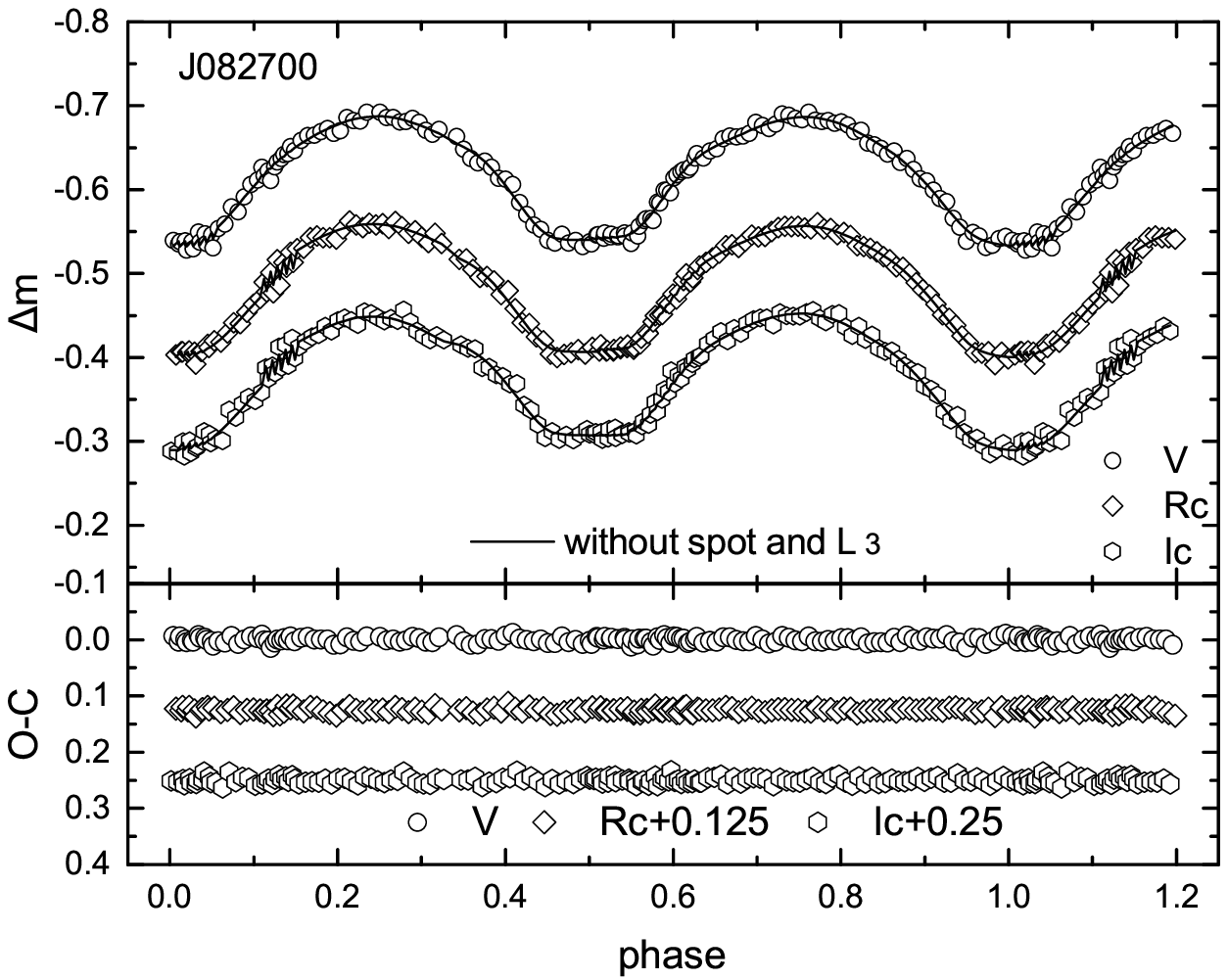}%
\includegraphics[width=0.32\textwidth]{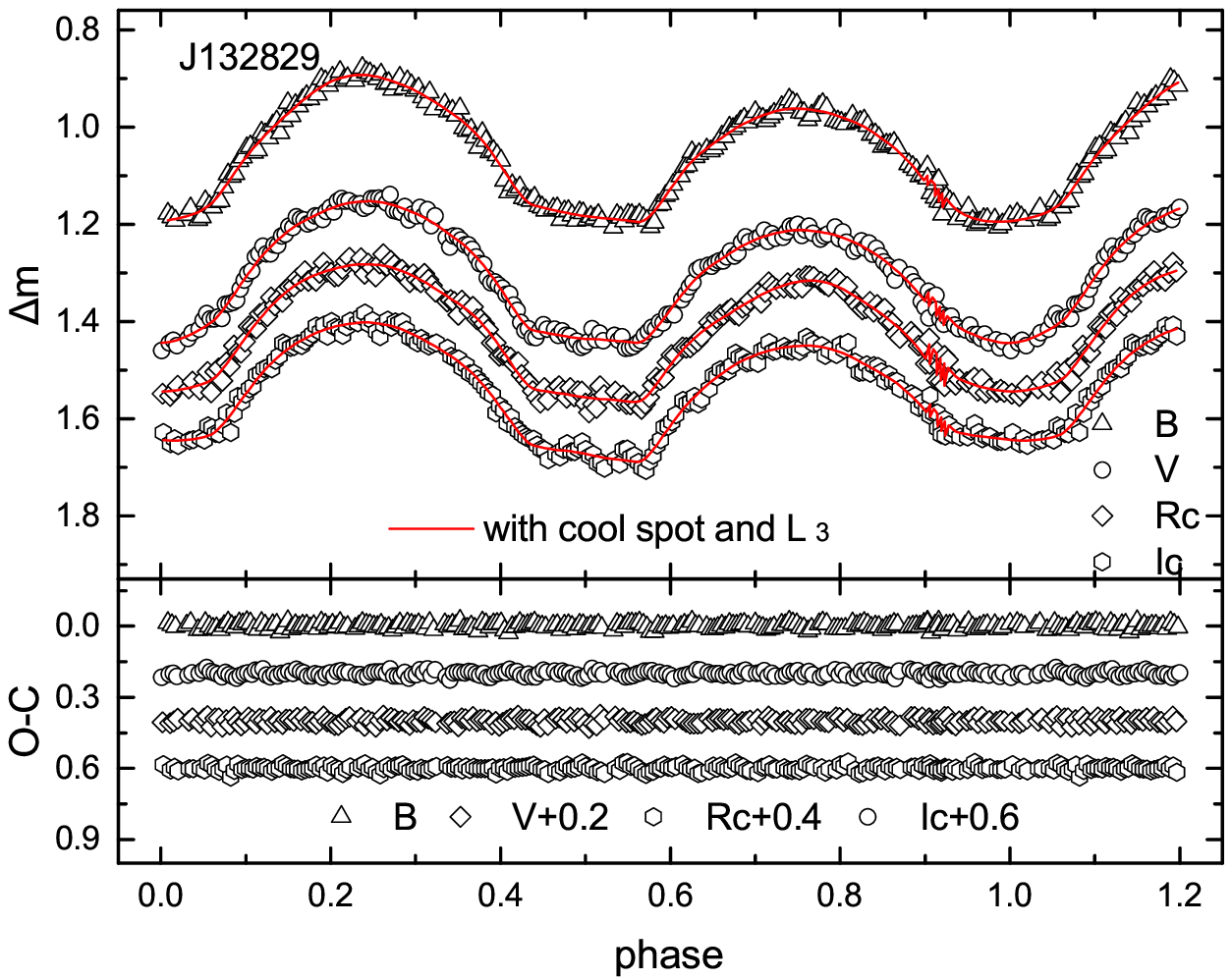}
\includegraphics[width=0.32\textwidth]{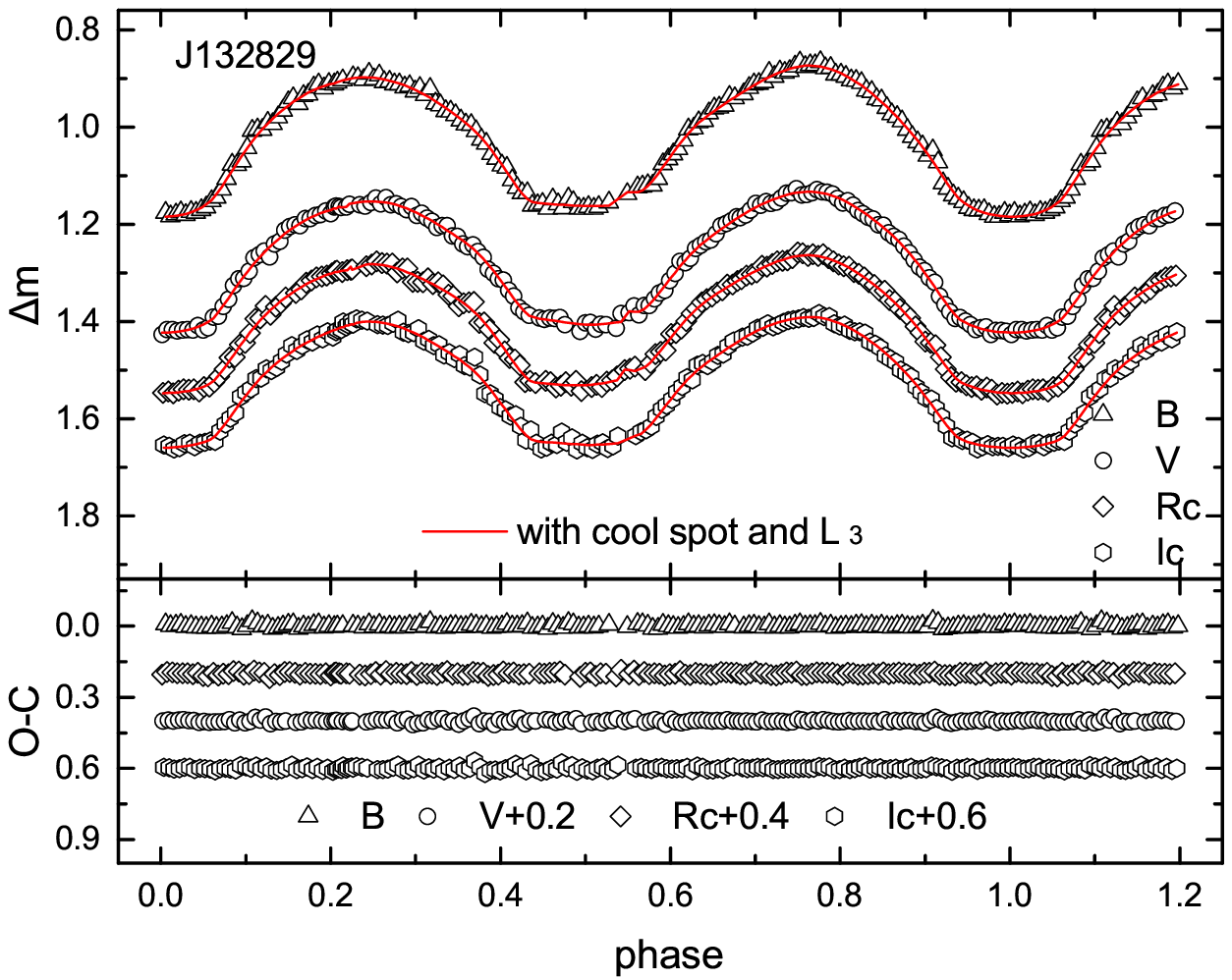}
\caption{The left panel displays the fitted light curves of J082700, the middle panel plots the fitted NAOs85cm light curve of J132829, while the right one plots the fitted WHOT light curve of J132829. The O-C residuals are plotted in the bottom panels, nearly flat residuals reveal that the theoretical light curves fit the observed ones very well. Different symbols represent different filters.}
\end{center}
\end{figure*}
\renewcommand\arraystretch{1.3}
\begin{table*}
\scriptsize
\begin{center}
\caption{Physical parameters of J082700 and J132829}
\begin{tabular}{llll}
\hline
\hline
Star  &  J082700   & J132829-NAOs85cm  &   J132829-WHOT       \\\hline
$T_1(K)$            &   5870$_{-332}^{+335}$          &    6275$_{-336}^{+337}$         &     6300$_{-339}^{+339}$            \\
$q(M_2/M_1) $       &   0.0550$_{-0.0004}^{+0.0006}$  &    0.0858$_{-0.0011}^{+0.0011}$ &     0.0889$_{-0.0011}^{+0.0011}$   \\
$T_2(K)$               &   5828$_{-405}^{+406}$          &    6375$_{-367}^{+369}$         &     6319$_{-366}^{+366}$            \\
$i(deg)$            &   68.7$_{-0.6}^{+0.6}$          &    81.4$_{-0.5}^{+0.5}$         &     81.5$_{-0.9}^{+0.9}$          \\
$\Omega_1=\Omega_2$ &   1.801$_{-0.002}^{+0.002}$     &    1.873$_{-0.003}^{+0.003}$    &     1.883$_{-0.004}^{+0.004}$     \\
$L_{2B}/L_{1B}$     &   -                             &    0.1395$_{-0.0040}^{+0.0044}$ &     0.1322$_{-0.0042}^{+0.0042}$ \\
$L_{2V}/L_{1V}$     &   0.0763$_{-0.0047}^{+0.0047}$  &    0.1354$_{-0.0033}^{+0.0035}$ &     0.1310$_{-0.0035}^{+0.0035}$ \\
$L_{2R_c}/L_{1R_c}$ &   0.0766$_{-0.0040}^{+0.0040}$  &    0.1337$_{-0.0030}^{+0.0032}$ &     0.1304$_{-0.0032}^{+0.0032}$ \\
$L_{2I_c}/L_{1I_c}$ &   0.0769$_{-0.0034}^{+0.0034}$  &    0.1320$_{-0.0027}^{+0.0029}$ &     0.1300$_{-0.0030}^{+0.0030}$ \\
$L_{3B}/L_{B}$      &   $-$                           &    0.0479$_{-0.0035}^{+0.0034}$ &     0.0293$_{-0.0077}^{+0.0079}$ \\
$L_{3V}/L_{V}$      &   $-$                           &    0.0191$_{-0.0039}^{+0.0039}$ &     0.0194$_{-0.0056}^{+0.0053}$ \\
$L_{3R_c}/L_{R_c}$  &   $-$                           &    0.0101$_{-0.0039}^{+0.0040}$ &     0.0112$_{-0.0046}^{+0.0046}$ \\
$L_{3I_c}/L_{I_c}$  &   $-$                           &    0.0080$_{-0.0004}^{+0.0004}$ &     0.0073$_{-0.0034}^{+0.0035}$ \\
$r_1$               &   0.6278$_{-0.0008}^{+0.0011}$  &    0.6113$_{-0.0011}^{+0.0011}$ &     0.6081$_{-0.0017}^{+0.0017}$   \\
$r_2$               &   0.1766$_{-0.0011}^{+0.0021}$  &    0.2209$_{-0.0023}^{+0.0024}$ &     0.2222$_{-0.0029}^{+0.0031}$    \\
$f$                 &   19$_{-5}^{+6}$ \%             &    74$_{-5}^{+5}$ \%            &     70$_{-7}^{+7}$ \%             \\
Spot                &   $-$                           &    Star 1                       &     Star 1                     \\
$\theta(deg)$       &   $-$                           &    98$_{-6}^{+6}$               &     38$_{-6}^{+6}$               \\
$\lambda(deg)$      &   $-$                           &    97$_{-4}^{+4}$               &     256$_{-12}^{+12}$               \\
$r_s(deg)$          &   $-$                           &    20$_{-1}^{+1}$               &     12$_{-4}^{+4}$               \\
$T_s$               &   $-$                           &    0.87$_{-0.03}^{+0.03}$       &     0.87$_{-0.16}^{+0.16}$       \\
\hline
\hline
\end{tabular}
\end{center}
\end{table*}

\section{Orbital Period Investigation}
Orbital period investigation is a very powerful tool to study the dynamical evolution of a binary star itself as well as the additional companion(s) orbiting the binary star (e.g., \citealt{Zhou2016b,Zhao2019,Liao2019,Li2019a,Pi2019}). Therefore, we analyzed the eclipsing time variations of J082700 and J132829. At present, there are many optical surveys that have observed these two targets. Northern Sky Variability Survey (NSVS, \citealt{Wozniak2004}), Catalina Sky Survey (CSS, \citealt{Drake2014}), and ASAS-SN have observed J082700, and NSVS, CSS, TAROT, WASP (\citealt{Butters2010}), and ASAS-SN have observed J132829. However, the time resolution of these surveys is very poor. In order to derive as many eclipse timings as possible, we construct the complete phase light curve or half of the phase light curve over longer time interval. One season of observations (several hundreds of data points) can be used because we are trying to identify the period variation of many years long, not the short-periodic signal shorter than several years. Figure 6 shows an example that we calculated the eclipsing times for J082700 by using the CSS data. We divided the data into three segments shown in the upper panel. For each segment, we shifted the data into one period using the equation, $MJD=MJD_0+P*E$, where $MJD$ is the observational time of the data, $MJD_0$ is the selected reference time, $P$ is the orbital period, and $E$ is the cycle number. Then, we derived the lower panel of Figure 6. Based on the light curves of the three segments displayed in the lower panel, five eclipsing minima were determined by the K-W method. Following this procedure, we determined two eclipsing minima by using NSVS data, five eclipsing minima by using CSS data, four eclipsing minima by using ASAS-SN data for J082700, and two eclipsing minima by using NSVS data, two eclipsing minima by using TAROT data, four eclipsing minima by using ASAS-SN data for J132829. The NSVS and CSS times were firstly transferred from MJD to JD ((MJD = JD - 2400000.5), and then to HJD. G. Srdoc firstly identified J082700 as a binary, and uploaded the figure of its light curve to VSX. Then, we calculated two eclipsing minima from the light curves\footnote{We communicated with G. Srdoc to ask for the observational data by email. However, he told us that the data were lost due to backup disc failure. So, we derived the original data points from the VSX figures by using GetData Graph Digitizer version 2.25.0.32 (http://getdata-graph-digitizer.com/). The times were firstly transferred from UTC to JD, and then to HJD.}. For J132829, we can derive one eclipsing minimum from the original WASP data, and no eclipsing minimum can be determined by using the CSS data. All the derived times of eclipsing minimum are listed in Table 2.

Using the following two equations,
\begin{equation}
\textrm{Min.I}=2455544.64369+0.2771582\textrm{E},
\end{equation}
\begin{equation}
\textrm{Min.I}=2454604.41693+0.384705\textrm{E},
\end{equation}
we constructed the $O-C$ diagram for J082700 and J132829, respectively. The corresponding curves are shown in the left and right panels of Figure 7. As seen in the left panel of Figure 7, the $O-C$ curve of J082700 exhibits a downward parabola plus a qusi-cyclic variation, so we used a second-order polynomial plus the light travel time effect model proposed by \cite{Irwin1952} to fit the curve,
\begin{eqnarray}
O-C&=& \Delta T_{0} + \Delta P_{0}\times E+{\beta \over 2}\times E^2 + K{1\over\sqrt{1-e^2\cos^2\omega} }[(1-e^{2})\frac{\sin(\nu+\omega)}{1+e\cos\nu}+e\sin\omega],
\end{eqnarray}
where $\Delta T_{0}$ and $\Delta P_{0}$ are respectively
the corrections of the initial epoch and period with respect to the values in Equation (2), $\beta$ is the long-term changing rate of the orbital period, the details of other parameters can be found in \cite{Irwin1952}.
As seen in the right panel of Figure 7, the $O-C$ curve of J132829 exhibits a downward parabola, the following second-order polynomial was applied to fit the curve,
\begin{eqnarray}
O-C= \Delta T_{0} + \Delta P_{0}\times E+{\beta \over 2}E^2.
\end{eqnarray}
The derived parameters are all listed in Table 3. We found that both of the two binaries show long-term orbital period decrease, and removing the long-term decrease, J082700 exhibits a cyclic variation with a period of 12.5 years and an amplitude of 0.0169 days. Because of the short time span and the scarcity of the eclipsing times of the two binaries, these results are required more observations to be confirmed.

\begin{figure*}
\begin{center}
\includegraphics[width=0.6\textwidth]{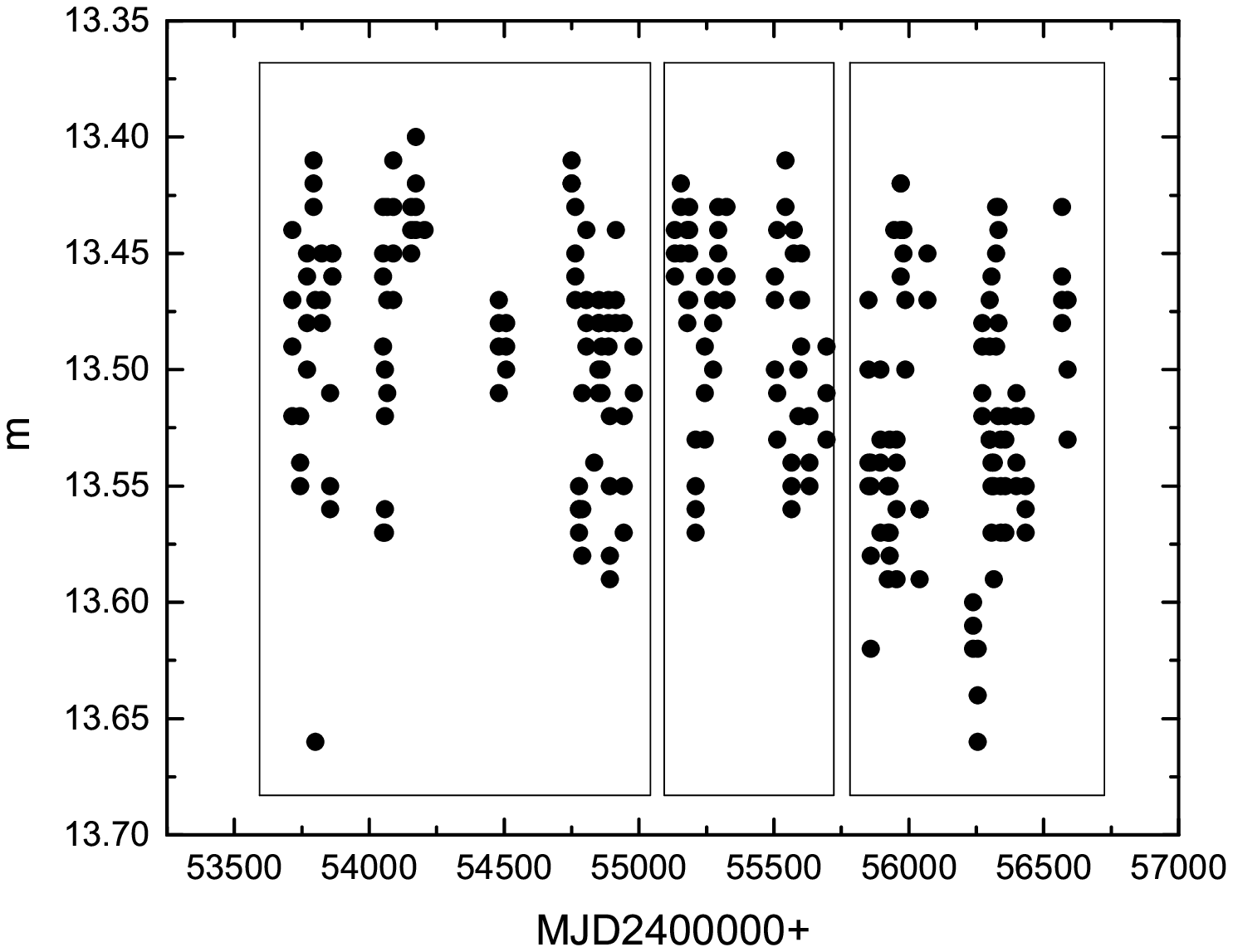}\\
\includegraphics[width=0.32\textwidth]{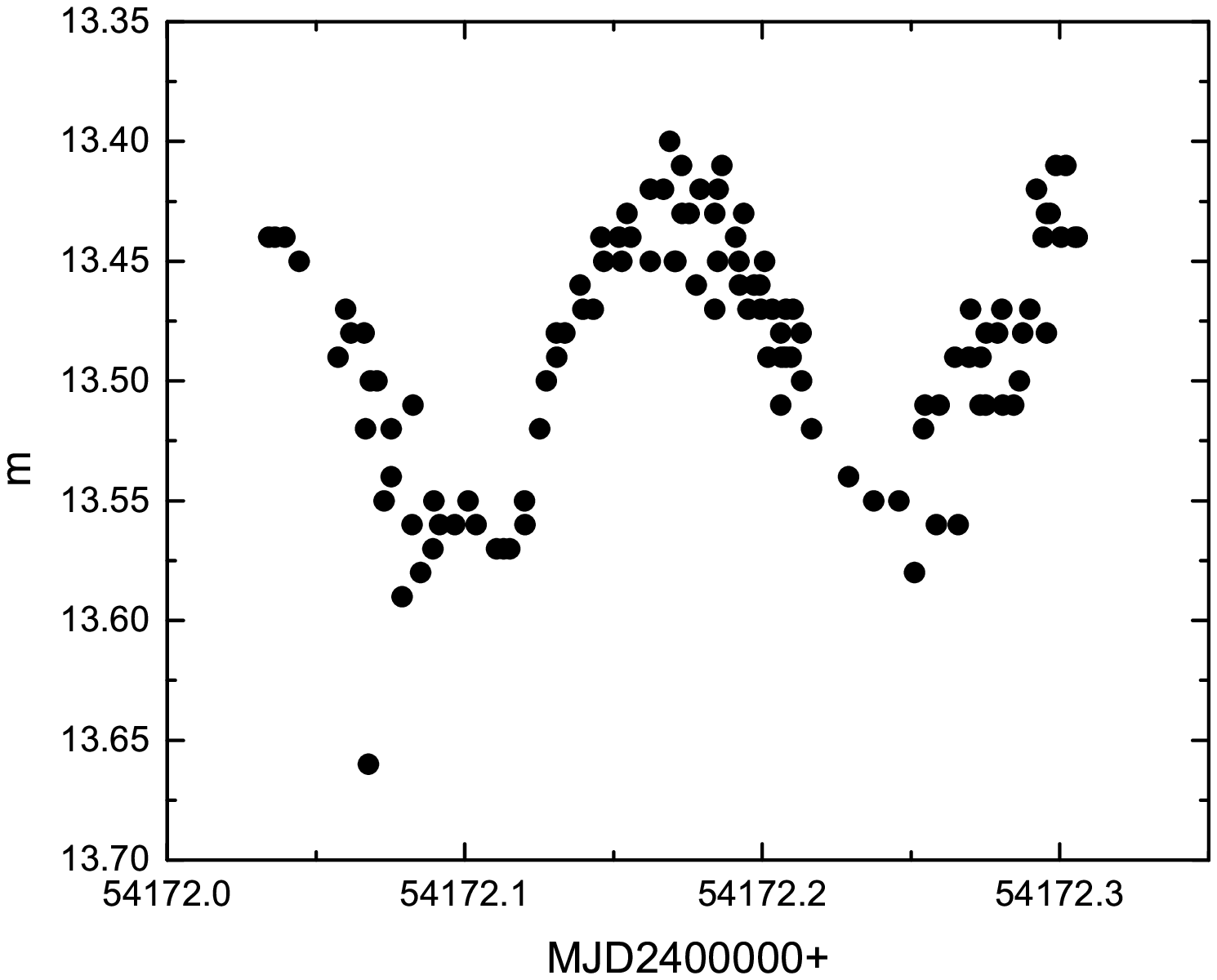}
\includegraphics[width=0.32\textwidth]{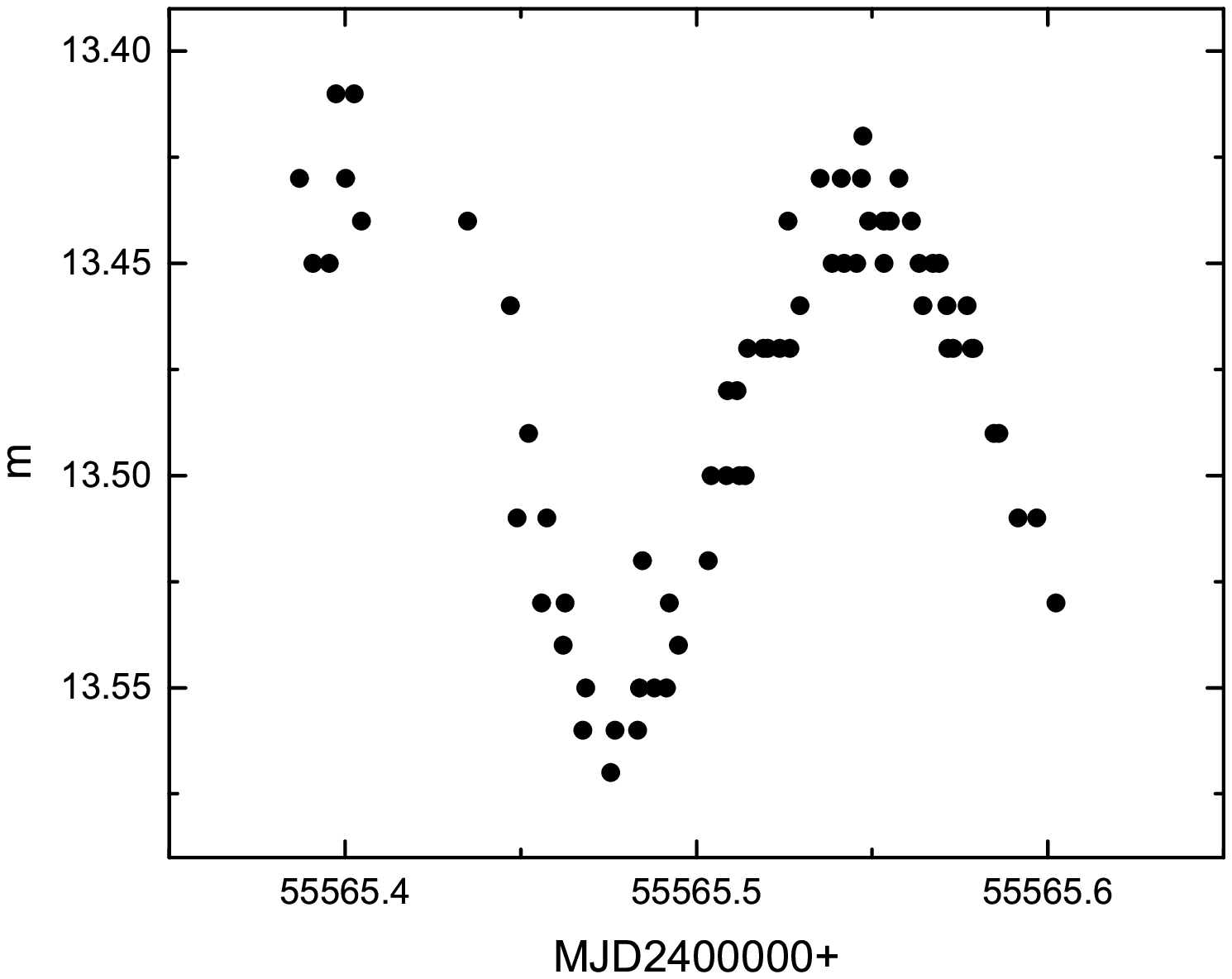}
\includegraphics[width=0.335\textwidth]{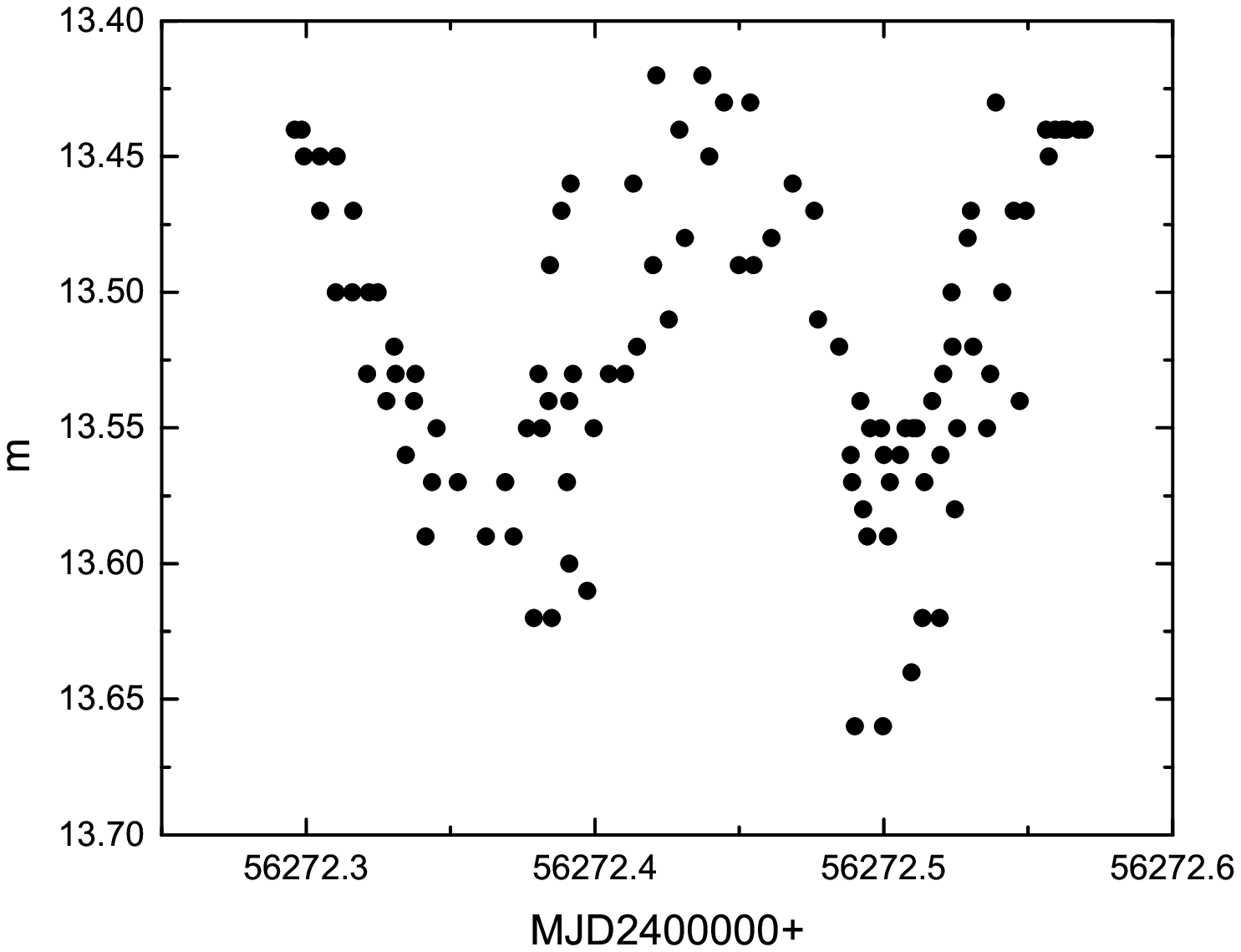}
\caption{The upper panel displays the light curves of J082700 observed by CSS, we divided the data into three segments and showed them in the lower panel. For each segment, we shifted the data into one period using the equation, $MJD=MJD_0+P*E$, where $MJD$ is the observational time of the data, $MJD_0$ is the selected reference time, $P$ is the orbital period, and $E$ is the cycle number.}
\end{center}
\end{figure*}

\begin{table}
\tiny
\begin{center}
\caption{Eclipsing Times of J082700 and J132829}
\label{Tab:ecl$-$times}
\begin{tabular}{cccccc|cccccc}
\hline
              \multicolumn{ 6}{c}{J082700} &        \multicolumn{ 6}{c}{ J132829}       \\\hline
HJD         &  Errors &  E        & O$-$C     &  Residuals& References& HJD        & Errors & E       & O$-$C     & Residuals& References\\
2400000+    &         &           &         &           &           & 2400000+   &        &         &         &          &           \\\hline
51555.63771   &  0.00252  &  $-$14392.5   & $-$0.00659  &  $-$0.0010  & (1)    & 51403.61957   & 0.00312  & $-$8320   & $-$0.05176 &0.00254    & (1)           \\
51555.76829   &  0.00186  &  $-$14392     & $-$0.01459  &  $-$0.0029  & (1)    & 51403.81857   & 0.00245  & $-$8319.5 & $-$0.04511 &0.00919    & (1)    \\
54172.60088   &  0.00102  &  $-$4950.5    & 0.02886   &  0.0006   & (2)    & 53644.17074   & 0.00143  & $-$2496   & $-$0.02251 &$-$0.00666   & (6)  \\
54172.74684   &  0.00122  &  $-$4950      & 0.03624   &  0.0006   & (2)    & 53644.35992   & 0.00208  & $-$2495.5 & $-$0.02568 &$-$0.00984   & (6)  \\
55544.50184   &  0.00060  &  $-$0.5       & $-$0.00327  &  0.0011   & (3)    & 54604.41693   & 0.00153  & 0       & 0        &0.00424    & (7)    \\
55544.64369   &  0.00041  &  0          & 0         &  0.0009   & (3)    & 56830.71560   & 0.00134  & 5787    & 0.01084  &$-$0.00056   & (4)  \\
55558.50158   &  0.00029  &  50         & $-$0.00002  &  0.0005   & (3)    & 56830.91283   & 0.00091  & 5787.5  & 0.01571  &0.00432    & (4)  \\
55558.63446   &  0.00058  &  50.5       & $-$0.00572  &  0.0007   & (3)    & 57766.89689   & 0.00133  & 8220.5  & 0.01251  &$-$0.00076   & (4) \\
55565.98239   &  0.00075  &  77         & $-$0.00248  &  0.0004   & (2)    & 57767.08849   & 0.00110  & 8221    & 0.01176  &$-$0.00152   & (4) \\
56272.86762   &  0.00259  &  2627.5     & $-$0.00924  &  0.0002   & (2)    & 58558.04146   & 0.00089  & 10277   & 0.01125  &$-$0.00144   & (8)\\
56273.00591   &  0.00330  &  2628       & $-$0.00953  &  0.0017   & (2)    & 58558.23706   & 0.00088  & 10277.5 & 0.01449  &0.00181    & (8)\\
57108.76828   &  0.00067  &  5643.5     & $-$0.01771  &  0.0007   & (4)    & 58599.20602   & 0.00036  & 10384   & 0.01237  &$-$0.00023   & (5)    \\
57108.90714   &  0.00066  &  5644       & $-$0.01743  &  0.0009   & (4)    &               &          &         &          &           & \\
58034.05724   &  0.00100  &  8982       & $-$0.02140  &  0.0005   & (4)    &               &          &         &          &           & \\
58034.19866   &  0.00130  &  8982.5     & $-$0.01856  &  0.0011   & (4)    &               &          &         &          &           & \\
58574.10494   &  0.00055  &  10930.5    & $-$0.01646  &  0.0006   & (5)    &               &          &         &          &           & \\
\hline
\end{tabular}
\end{center}
\scriptsize
(1) NSVS; (2) CSS; (3) VSX; (4) ASAS$-$SN; (5) WHOT; (6) TAROT; (7) WASP; (8) NAOs85cm. \\
\end{table}

\begin{figure*}
\begin{center}
\includegraphics[width=0.465\textwidth]{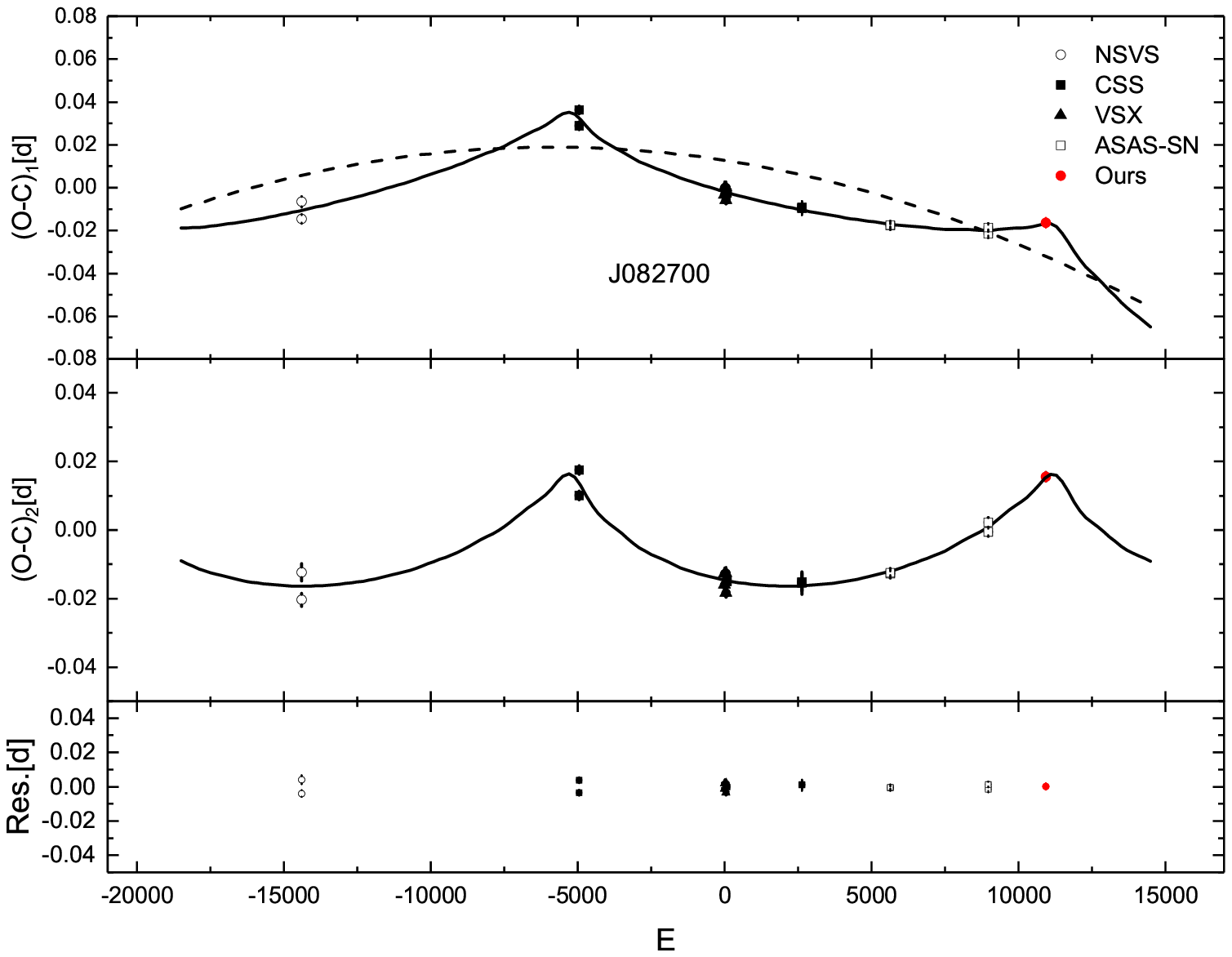}%
\includegraphics[width=0.5\textwidth]{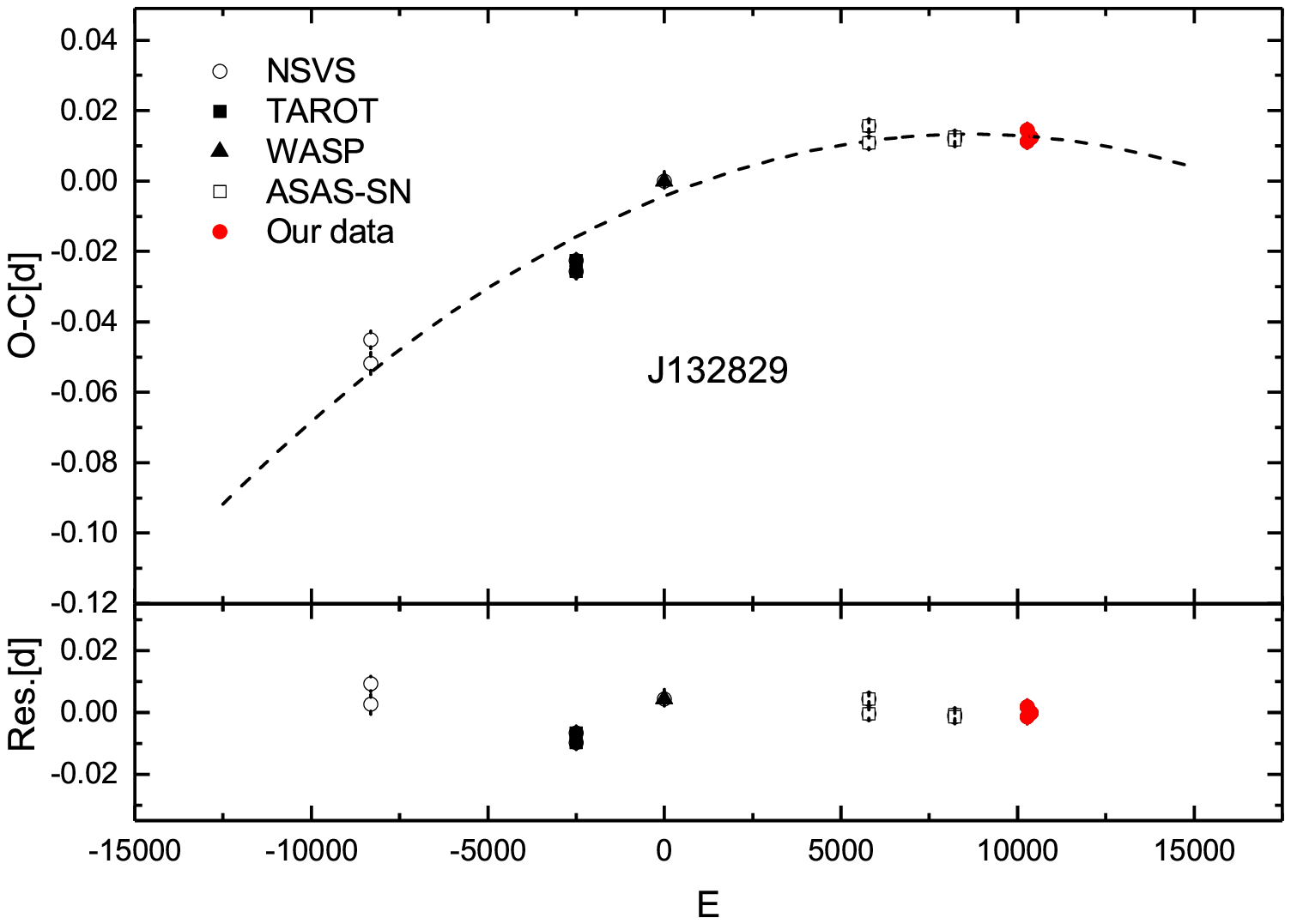}

\caption{The left panel shows the $O-C$ diagram of J082700,  while the right one plots that of J132829. Different symbols refer to different observations.}
\end{center}
\end{figure*}

\begin{table}
\begin{center}
\caption{Parameters determined by Equations (3) and (4) for J082700 and J132829, respectively}
\begin{tabular}{lcc}
\hline\hline
Parameters &     J082700 &     J132829 \\
\hline
$\Delta T_0$ (days) &    0.01271$(\pm0.00124)$ &  $-0.00424(\pm0.00230)$  \\

$\Delta P_0$ (days) & $-2.12(\pm0.93)\times10^{-6}$ & $4.06(\pm0.36)\times10^{-6}$  \\

$\beta$ (days $yr^{-1}$) & $-9.52(\pm4.01)\times10^{-7}$ & $-4.46(\pm0.81)\times10^{-7}$   \\

$K$ (d) &     0.0169($\pm0.0059$) & $-$    \\

$e$ &       0.82($\pm0.37$) &   $-$    \\

$P_{3}$ (yr) &     12.5($\pm3.2$) &   $-$   \\

$\omega (^\circ)$ &       108.7($ \pm47.5$) &  $-$    \\

$ T_P$ (HJD) &  2455886.2($\pm817.5$) &  $-$    \\
\hline
\end{tabular}
\end{center}
\end{table}

\section{Discussions and conclusions}
Using the MCMC parameter search method based on PHOEBE 2.3, we analyzed one set of complete multiple color light curves of J087200 and two sets of complete multiple color light curves of J132829. The results show that these two binaries are both ELMRCBs, $q\sim0.055$ for J082700 and $q\sim0.089$ for J132829. J082700 is a shallow contact binary ($f\sim19\%<25\%$), while J132829 is a deep contact binary ($f\sim70\%>50\%$). The asymmetric light curves of J132829 can be interpreted by a dark spot on the more massive component. As seen in Figure 1, both of the light curves of J082700 and J132829 show clearly total eclipsing secondary minima. As proposed by \cite{Pribulla2003} and \cite{Terrell2005}, the physical parameters of the two systems determined only by photometric light-curve synthetics should be reliable. We should state that none of the current models of contact binaries (W-D, PHOEBE,...) account for energy transfer in the envelope correctly, yet at such low mass ratios that has to be extremely significant. In addition, the fillout factor at such low mass ratios is poorly defined because surface potentials through the inner Lagrange point ($L_1$) and the outer Lagrange point ($L_2$) are numerically very close.
By calculating eclipsing times from all available photometric surveys and observations, we investigated the $O-C$ variations of the two binaries. The orbital period of J082700 displays a long-term decrease at a rate of $-9.52\times10^{-7}$ d yr$^{-1}$ plus a cyclic oscillation with a period of 12.5 years and an amplitude of 0.0169 days, while that of J132829 shows a long-term decrease at a rate of $-4.46\times10^{-7}$ d yr$^{-1}$.

\subsection{The period changes}
Both of the two systems exhibit long-term period decrease. Such changes can be generally caused by the mass transfer from the more massive component to the less massive one or by the angular momentum loss (AML) or the combination. To discuss the reason of the long-term period decrease, we have to derive the absolute parameters of the two binaries. However, we cannot directly determine their absolute parameters because of the lack of the radial velocity curve. Nevertheless, we try to estimate their global parameters based on the final results of the light curve solutions. By assuming the more massive primary components of the two binaries are main-sequence stars, the masses can be estimated to be $M_1=1.06M_\odot$ for J082700 and $M_1=1.23M_\odot$ for J132829 based on the online Table\nolinebreak{http://www.pas.rochester.edu/\textasciitilde emamajek/EEM\_dwarf\_UBVIJHK\_colors\_Teff.txt} \citep{Pecaut2013}. According to the adopted photometric elements and the Kepler's third law, the global parameters of the two binaries can be estimated as follows: $a=1.83R_\odot$, $M_1=1.06M_\odot$, $M_2=0.06M_\odot$, $R_1=1.15R_\odot$, $R_2=0.32R_\odot$, $L_1=1.40L_\odot$ and $L_2=0.11L_\odot$ for J082700, and $a=2.45R_\odot$, $M_1=1.23M_\odot$, $M_2=0.11M_\odot$, $R_1=1.49R_\odot$, $R_2=0.55R_\odot$, $L_1=3.15L_\odot$ and $L_2=0.43L_\odot$ for J132829.
If the period decrease is caused by the mass transfer, the following equation can be used to estimate the mass transfer rate,
\begin{eqnarray}
{\dot{P}\over P}=-3\dot{M_1}({1\over M_1}-{1\over M_2}) .
\end{eqnarray}
The mass transfer rate was calculated to be $dM_1/dt=-7.06(\pm2.98)\times10^{-8}\,M_\odot$ yr$^{-1}$ for J082700 and $dM_1/dt=-4.64(\pm0.85)\times10^{-9}\,M_\odot$ yr$^{-1}$ for J132829. The negative sign reveals that the more massive primary is losing mass. Assuming constant angular momentum and the total mass, the thermal timescale can be calculated to be $\tau_{th}=2.20\times10^7$ years for J082700 and $\tau_{th}=1.01\times10^7$ years for J132829 using the equation $\tau_{th}={GM_1^2\over R_1L_1}$. The thermal timescale mass transfer rate is determined to be $M_1/\tau_{th}=4.82\times10^{-8}\,M_\odot$ yr$^{-1}$ for J082700, while that for J132829 is derived to be $M_1/\tau_{th}=1.22\times10^{-7}\,M_\odot$ yr$^{-1}$. For J082700, the thermal mass transfer rate is similar with that calculated by Equation (6), meaning that the long-term orbital period decrease is possibly caused by mass transfer. However, we cannot rule out the probability of AML via magnetic stellar wind. For J132829, the thermal mass transfer rate is very different from that calculated by Equation (6), meaning that the long-term orbital period decrease is possibly due to AML.

The cyclic oscillation in the $O-C$ diagram of J082700 can be generally resulted from the magnetic activity of one or two components (\citealt{Applegate1992}) or the light travel time effect (LTTE) due to a third body (e.g., \citealt{Liao2010,Lee2013,Zhou2017}). In order to check which one is the more likely reason, we calculated the quadruple moment variations of the two components using the equations taken from \cite{Applegate1992} and \cite{Lanza2002},
\begin{eqnarray}
{\Delta P\over P}={2\pi A\over P_{mod}} ,
\end{eqnarray}
\begin{eqnarray}
{\bigtriangleup P\over P}=-9{\Delta Q\over {Ma^2}} .
\end{eqnarray}
$\Delta Q_1=1.77\times10^{50}$ g cm$^2$ and $\Delta Q_2=9.72\times10^{48}$ g cm$^2$ were determined respectively for the primary and secondary components. The value of quadruple moment variation of the primary component is close to the typical value of $10^{51}-10^{52}$ g cm$^2$. Therefore, Applegate mechanism is a possible reason for the cyclic oscillation. Although Applegate mechanism can explain the cyclic oscillation, we cannot rule out the possibility of LTTE via a third body. So we calculated the mass function of the third body using the following equation,
\begin{equation}
f(m)={(m_{3}\sin i)^3\over (m_1+m_2+m_{3})^2}={4\pi\over GP^2_{3}}\times(a_{12}\sin i)^3,
\end{equation}
and determined $f(m)=1.61(\pm1.68)\times10^{-1}\,M_\odot$. If the third body is coplanar with the central eclipsing pair ($i^\prime=i=66^\circ.9$), the mass and the separation were derived to be $M_3=0.96\pm0.75\,M_\odot$ and $a_3=4.27\pm3.65$ AU. Considering the mass of the tertiary companion, the luminosity would be $0.8\,L_\odot$ for a main-sequence star. However, when we searched third light during the light curve modeling, no third light was detected. Maybe the tertiary companion is a very faint star or a compact object.

\subsection{Statistics on ELMRCBs}
In order to investigate the properties of ELMRCBs, we carried out a statistics analysis on contact binaries with mass ratios $q\lesssim0.1$. The name, period, mass ratio, inclination, the effective temperatures of the two components, the contact degree, the orbital period change rate, and the spin angular momentum to the orbital angular momentum ratio, $J_{spin}/J_{orb}$, are shown in Table 4. Sixteen such systems were collected. In this table, $J_{spin}/J_{orb}$ was calculated using the following equation taken from \cite{Yang2015},
\begin{equation}
{J_{spin}\over J_{orb}}={{1+q}\over q}[(k_1r_1)^2+(k_2r_2)^2q],
\end{equation}
where $r_{1,2}$ and $k_{1,2}$ are the relative radii and the gyration radii to the stellar radii ratio. For W UMa type contact binaries, $k_1^2=k_2^2=k^2$, and taking $k^2=0.06$ from \cite{Li2006}, the values of $J_{spin}/J_{orb}$ were derived. As seen in this table, the mass ratios of two stars (V1187 Her and J082700) are below the theoretically predicted cut-off mass ratio of contact binaries. And we found that the value of $J_{spin}/J_{orb}$ has a tendency to decrease with the reduction of the mass ratio.

A very confusing result was obtained from this statistics, the values of $J_{spin}/J_{orb}$ of three targets are greater than 1/3 (V1187 Her, J082700, and V857 Her). According to \cite{Hut1980}, binary systems will be dynamically unstable when the spin angular momentum is more than one third of the orbital angular momentum. These three targets should be dynamically unstable. However, researchers have observed stable light variability of eclipse, and their orbital period changes show no abnormal behavior. These three stars are stable contact binaries. There are two possible reasons. The first is that the dynamical stability criterion proposed by \cite{Hut1980} should be corrected because some physical processes that are not known to date may not have been taken into account. If considering the unknown physical processes, it will possibly lead to the change of the present dynamical stability limit (the value of $J_{spin}/J_{orb}$ would be greater than 1/3).
The other reason, most likely, is that the ratio of the gyration radii to the stellar radii ($k$) should be smaller than $k^2=0.06$, meaning that the mass ratio limit of contact binaries will be affected by this ratio. The smaller the $k$, the smaller the minimum mass ratio of contact binaries. This is consistent with the results derived by \cite{Rasio1995} and \cite{Jiang2010}.
If V1187 Her is just at the boundary of stability, $k^2=0.03249$ was determined. Then, the revised values of $J_{spin}/J_{orb}$ are listed in the tenth column of Table 4 by using $k^2=0.03249$. The correlations between the mass ratio and the initial and revised $J_{spin}/J_{orb}$ are respectively displayed in the left and right panels of Figure 8. The dashed line represents the boundary between stable and unstable. A least-square method yields the following equations,
\begin{equation}
{J_{spin}\over J_{orb}}=0.20(\pm0.02)+2.57(\pm0.63)\times e^{-41.72(\pm6.04)\times q},
\end{equation}
for the left panel, and
\begin{equation}
{J_{spin}\over J_{orb}}=0.11(\pm0.01)+1.39(\pm0.34)\times e^{-41.72(\pm6.04)\times q},
\end{equation}
for the right panel.
Based on Equation (11), a predicted minimum mass ratio was derived to be 0.0716, which is corresponding to the result derived by \cite{Li2006}. Due to Equation (12), a predicted cut-off mass ratio was determined to be 0.0439, which is consistent with the result determined by \cite{Yang2015}.

By the statistics, not all contact binaries with mass ratios $q\lesssim0.1$ are deep contact binaries, some of them are shallow or medium contact systems, even one of them has a fill-out factor of 1\% (NSVS 2569022). This may mean that the formation of ELMRCBs has two channels. One channel is for the shallow or medium contact systems, their initial short period detached binaries had extremely low mass ratios and formed ELMRCBs by AML, and these ELMRCBs are newly formed contact binaries. The other channel is for deep contact ones, they are likely to have experienced the process of the mass transfer from the less massive component to the more massive one, and now they are at the late evolutionary stage and losing mass from the outer Lagrange point.

\begin{figure}
\begin{center}
\includegraphics[angle=0,scale=0.5]{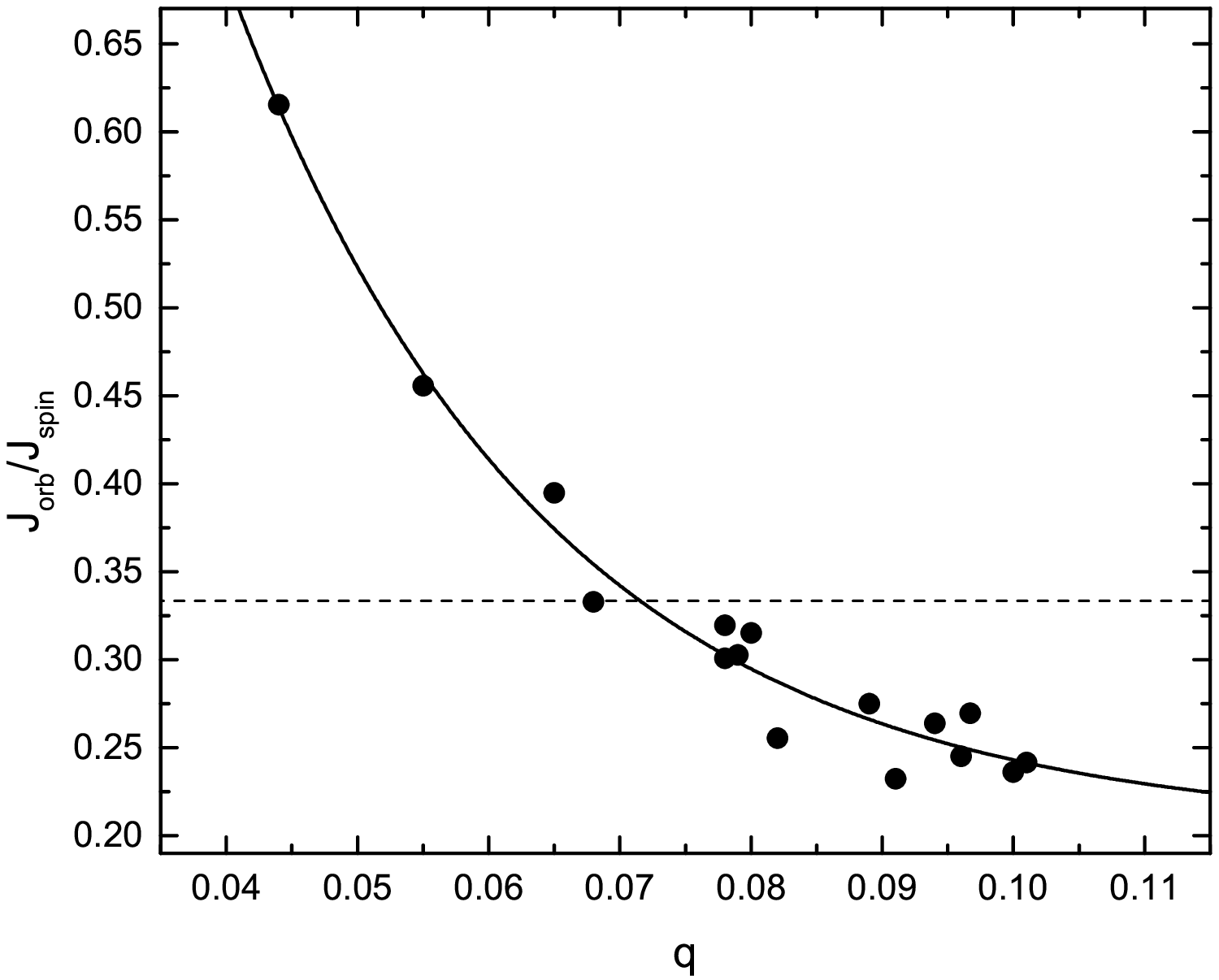}
\includegraphics[angle=0,scale=0.5]{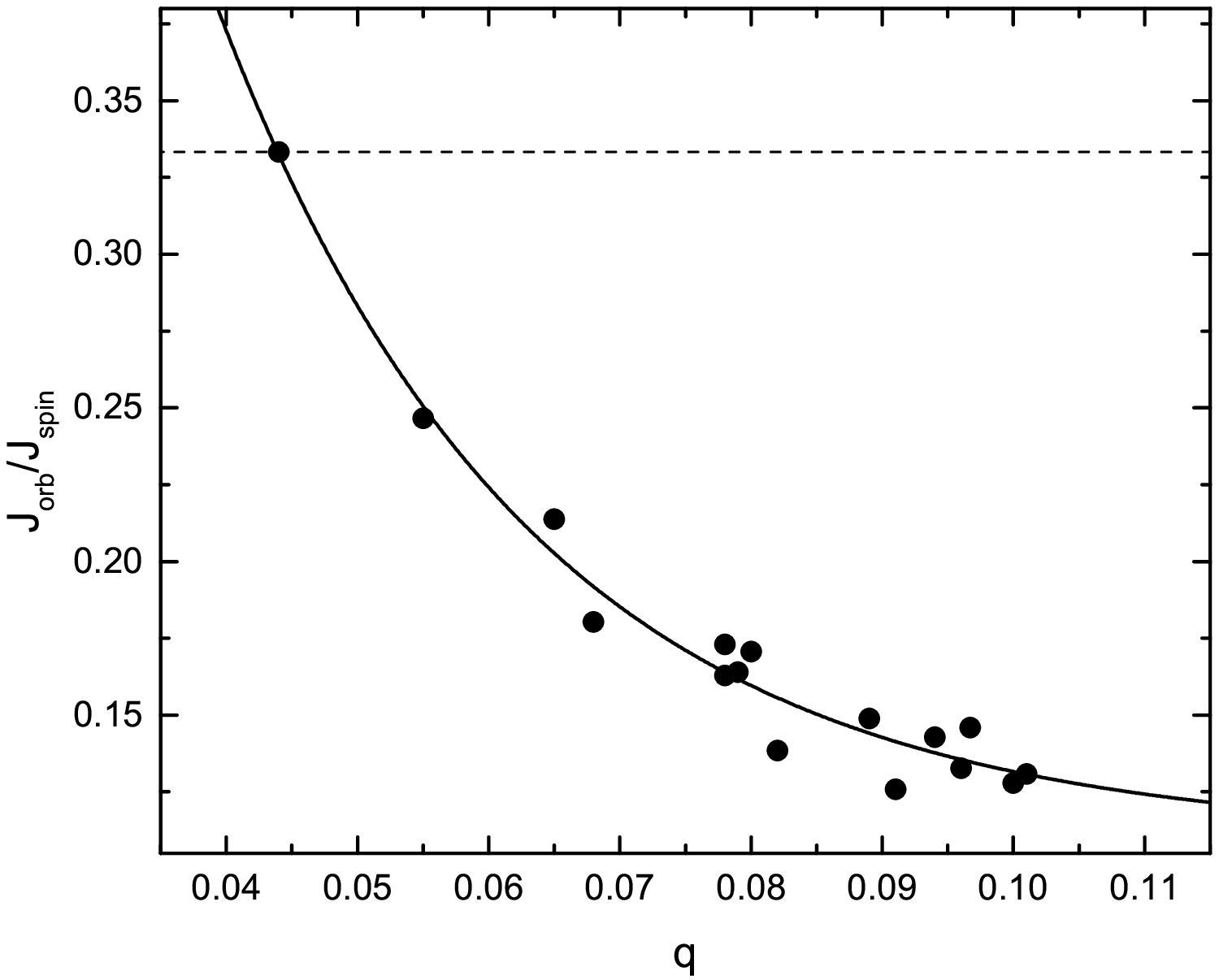}
\caption { $J_{spin}/J_{orb}$ versus mass ratio. Left panel shows the results when $k^2=0.06$, while the right one shows the results when $k^2=0.03249$. The dashed line represents dynamical stability limited. The solid circles denote the sixteen ELMRCBs.
The solid line refers to an exponential fit to the sixteen points.}
\end{center}
\end{figure}

\begin{table*}
\tiny
\begin{center}
\caption{Contact binaries with mass ratios $q\lesssim0.1$}
\begin{tabular}{llccccccccc}
\hline
      Parameters        &      Period(d) & $q$    &$i$(deg)&T$_1$(K) &T$_2$(K)&  $f$ & $dP/dt$(days yr$^{-1}$) &$J_{spin}/J_{orb}$&$J_{spin}/J_{orb}$$^c$  &References \\\hline
     V1187 Her$^*$      &       0.310764 &  0.044 & 66.7   &  6250   & 6680   & 84\% &  $-1.50\times10^{-7}$   &0.616    &0.333           &   (1)    \\
     J082700$^*$      &       0.277158 &  0.055 & 68.7   &  5870   & 5728   & 19\% &  $-9.52\times10^{-7}$   &0.456    &0.247           &   (2)    \\
      V857 Her$^*$      &       0.382230 &  0.065 & 85.4   &  8300   & 8513   & 84\% &  $+2.90\times10^{-7}$   &0.395    &0.214           &   (3)    \\
ASAS J083241+2332.4$^*$ &       0.311321 &  0.068 & 82.7   &  6300   & 6672   & 69\% &  $+8.80\times10^{-7}$   &0.333    &0.180           &   (4) \\
 NSVS 2569022$^*$       &       0.287797 &  0.078 & 76.3   &  6100   & 6100   &  1\% &  $-$                    &0.301    &0.163           &   (5)     \\
       M4 V53$^*$       &       0.308449 &  0.078 & 74.4   &  7415   & 6611   & 69\% &  $-5.89\times10^{-8}$   &0.320    &0.173           &   (6)     \\
       SX Crv$^\dagger$ &       0.316599 &  0.079 & 61.2   &  6340   & 6160   & 27\% &  $-1.05\times10^{-6}$   &0.303    &0.164           &   (7)    \\
     V870 Ara$^\dagger$ &       0.399722 &  0.082 & 70.0   &  5860   & 6210   & 96\% &  $-$                    &0.256    &0.138           &   (8) \\
       NSV 13890$^*$    &       0.373880 &  0.080 & 76.2   &  6510   & 6426   & 90\% &  $-$                    &0.315    &0.171           &   (9)        \\
       J132829$^*$      &       0.384705 &  0.089 & 81.5   &  6300   & 6319   & 70\% &  $-4.46\times10^{-8}$   &0.275    &0.149           &   (2)     \\
       KR Com$^\dagger$ &       0.407970 &  0.091 & 52.1   &  5549   & 6072   & shallow?&  $-$                 &0.232    &0.126           &   (10)      \\
       V1309 Sco$^*$    &       1.445600 &  0.094 & 73.4   &  4500   & 4354   & 89\% &  decreasing             &0.264    &0.143           &   (11)     \\
       FP Boo$^\dagger$ &       0.640482 &  0.096 & 68.8   &  6980   & 6456   & 38\% &  $-5.89\times10^{-8}$   &0.245    &0.133           &   (12)      \\
       KIC 11097678$^*$ &       0.999716 &  0.097 & 85.1   &  6493   & 6426   & 87\% &  $-$                    &0.270    &0.146           &   (13)      \\
       XX Sex$^\dagger$ &       0.540108 &  0.100 & 74.9   &  6881   & 6378   & 42\% &  $-$                    &0.236    &0.128           &   (14)        \\
       AW CrB$^*$       &       0.360935 &  0.101 & 82.1   &  6700   & 6808   & 75\% &  $+3.58\times10^{-7}$   &0.242    &0.131           &   (15)        \\
\hline
\end{tabular}
\end{center}
(1) \cite{Caton2019}; (2) This paper; (3) \cite{Qian2005b}; (4) \cite{Sriram2016}; (5) \cite{Kjurkchieva2018}; (6) \cite{Li2017};
(7) \cite{Zola2004}; (8) \cite{Szalai2007}; (9) \cite{Wadhwa2006}; (10 \cite{Zasche2010} (11) \cite{Zhu2016}; (12) \cite{Gazeas2006}; (13) \cite{Zola2017};
(14) \cite{Deb2011}; (15) \cite{Broens2013}.\\
$^*$ The mass ratios of these systems were determined photometrically.\\
$^\dagger$ The mass ratios of these systems were determined spectroscopically.\\
$^c$ The revised value of $J_{spin}/J_{orb}$.\\
Notes: The physical parameters of V1309 Sco were determined by the light curve before the merge.\\
The mass ratio of AW CrB was determined to be 0.099 when no spot was considered, so we added this target in our list.
\end{table*}

\subsection{The evolutionary states of these ELMRCBs}
Most of our samples listed in Table 4 have no radial velocity observations. We cannot determine their absolute parameters directly. In order to investigate their evolutionary states, we constructed the color-density diagram. The color of the two components of all the targets can be computed by interpolating their temperatures into the online Table\nolinebreak{http://www.pas.rochester.edu/\textasciitilde emamajek/EEM\_dwarf\_UBVIJHK\_colors\_Teff.txt} provided by \cite{Pecaut2013}, and the mean density can be calculated by using Equations (4a) and (4b) in \cite{Mochnacki1981}. Then, the color-density diagram is plotted in Figure 9. The zero age main sequence (ZAMS) and the terminal age main sequence (TAMS) indicated by solid and dashed lines were taken from \cite{Mochnacki1981}. In order to compare with normal contact binaries, we also plotted the low mass contact binaries (LMCBs) extracted from \cite{Yakut2005} in Figure 9. We can see that the evolutionary status of the components of ELMRCBs are similar with those of LMCBs. The more massive components are evolved stars because they are located between the ZAMS and TAMS lines or below TAMS line, while the less massive ones are non-evolved or little evolved stars because they are located around the ZAMS. For the secondary components of ELMRCBs, two (KIC 11097678 and V1309 Sco) of them are located very far below the TAMS line, meaning that they are more evolved stars. We found that these two binaries (KIC 11097678 and V1309 Sco) have the longest period among the ELMRCBs, this is the reason why their secondary components are located below TAMS line, they should be subgiants or red giants.

\begin{figure*}
\begin{center}
\includegraphics[angle=0,scale=0.7]{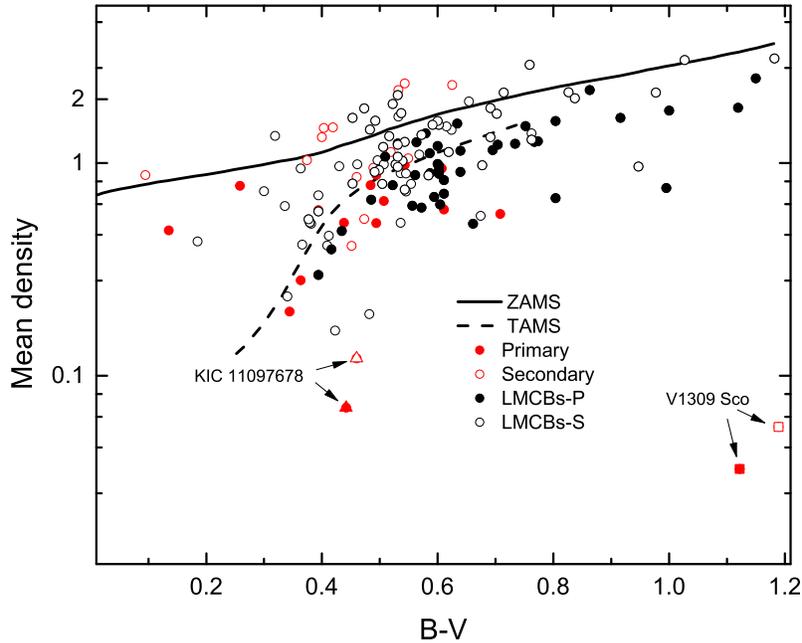}
\caption{This color-density diagram for ELMRCBs. Solid symbols show the more massive components, while the open symbols represent the less massive ones. The red symbols represent the ELMRCBs, while the black symbols refer to the LMCBs extracted from \cite{Yakut2005}. The ZAMS and TAMS lines taken from Mochnacki (1981) are respectively indicated by solid and dashed lines. "P" means the more massive component, while "S" means the less massive one.
}
\end{center}
\end{figure*}

In conclusion, multi-color light curves of two contact binaries, J082700 and J132829, have been analyzed. They are members of ELMRCBs. By analyzing the eclipsing times of the two binaries, we found that both of them show long-term decrease. For J082700, a cyclic modulation is superimposed on the long-term variation, which is more likely caused by the Applegate mechanism. By the statics of ELMRCBs, we found that the values of the spin angular momentum to the orbital angular momentum ratio of three systems are more than 1/3. Two possible reasons may be considered: One is the dynamical stability limited predicted by \cite{Hut1980} would be changed. The other is the dimensionless gyration radius should be less than $k^2=0.06$, meaning that the cut-off mass ratio of contact binaries depends on the dimensionless gyration radius. Two channels of the formation of ELMRCBs are presented. The evolutionary states of ELMRCBs are also discussed, we found that although they are very unusual, their locations on the color-density diagram are consistent with normal W UMa contact binaries. For J082700, the value of $J_{spin}/J_{orb}$ is more than or very close to $1/3$, it will be quickly merging. For J132829, the fillout factor is 70\%, and will be increased with the decreasing orbital period, a dynamical instability will be encountered when the surfaces of the components touch the outer critical Roche lobe, then J132829 will be merging to a single fast-rotation star. Therefore, the two contact binaries are both potential progenitors of the merger. Future observations, especially the high resolution spectroscopic observations, are needed to confirm their extremely low mass ratios and to determine their more precise absolute parameters.

~\\
Thanks the referee very much for the very help and useful comments to improve our manuscript a lot. This work is supported by the Joint Research Fund in Astronomy (No. U1931103) under cooperative agreement between National Natural Science Foundation of China (NSFC) and Chinese Academy of Sciences (CAS), and by NSFC (No. 11703016), and by the Natural Science Foundation of Shandong Province (Nos. ZR2014AQ019, JQ201702), and by Young Scholars Program of Shandong University, Weihai (Nos. 20820162003, 20820171006), and by the Chinese Academy of Sciences Interdisciplinary
Innovation Team, and by the Open Research Program of Key Laboratory for the Structure and Evolution of Celestial Objects (No. OP201704). Work by CHK was supported by the grant of National Research Foundation of Korea (2020R1A4A2002885). The calculations in this work were carried out at Supercomputing Center of Shandong University, Weihai.

We acknowledge the support of the staff of the Xinglong 85cm
telescope, and WHOT. This work was partially supported by the Open Project Program of the Key
Laboratory of Optical Astronomy, National Astronomical Observatories, Chinese
Academy of Sciences.

This work has made use of data from the European Space Agency (ESA) mission
{\it Gaia} (\url{https://www.cosmos.esa.int/gaia}), processed by the {\it Gaia}
Data Processing and Analysis Consortium (DPAC,
\url{https://www.cosmos.esa.int/web/gaia/dpac/consortium}). Funding for the DPAC
has been provided by national institutions, in particular the institutions
participating in the {\it Gaia} Multilateral Agreement.

This paper makes use of data from the DR1 of the WASP data (\citealt{Butters2010}) as provided by the WASP consortium,
and the computing and storage facilities at the CERIT Scientific Cloud, reg. no. CZ.1.05/3.2.00/08.0144
which is operated by Masaryk University, Czech Republic.


\end{CJK}
\end{document}